\documentclass[sn-nature]{sn-jnl}

\usepackage{subcaption}
\usepackage{graphicx}%
\usepackage{multirow}%
\usepackage{amsmath,amssymb,amsfonts}%
\usepackage{amsthm}%
\usepackage{mathrsfs}%
\usepackage[title]{appendix}%
\usepackage{xcolor}%
\usepackage{textcomp}%
\usepackage{manyfoot}%
\usepackage{booktabs}%
\usepackage{algorithm}%
\usepackage{algorithmicx}%
\usepackage{algpseudocode}%
\usepackage{listings}%
\usepackage{enumitem}
\usepackage{placeins}

\usepackage[numbers]{natbib}
\usepackage[sectionbib]{bibunits}
\defaultbibliographystyle{sn-nature}
\defaultbibliography{bibexport}

\hypersetup{colorlinks=true,linkcolor=blue,urlcolor=blue}
\usepackage{macros_alpha}
\usepackage{dcolumn}
\usepackage{booktabs}
\newcolumntype{C}{>{$}c<{$}}
\AtBeginDocument{
\heavyrulewidth=.08em
\lightrulewidth=.05em
\cmidrulewidth=.03em
\belowrulesep=.65ex
\belowbottomsep=0pt
\aboverulesep=.4ex
\abovetopsep=0pt
\cmidrulesep=\doublerulesep
\cmidrulekern=.5em
\defaultaddspace=.5em
}
\usepackage{multirow}
\usepackage{defs}

\theoremstyle{thmstyleone}%
\theoremstyle{thmstyletwo}%

\theoremstyle{thmstylethree}%

\makeatletter
\definecolor{Green}{rgb}{0,0.7,0}
\definecolor{Orange}{rgb}{0.8, 0.3, 0}

\newcounter{CNOTE}
\newcommand{\anote}[2]{%
  \setlength{\marginparwidth}{2.5cm}
  \addtocounter{CNOTE}{1}
  \reversemarginpar
  \ifmmode
    {\color{Orange}\ensuremath{{}^{\{\arabic{CNOTE}\}}}}
    {\color{Green}\text{#2}}
  \else%
    \begin{marginpar}
       {\color{Orange}\footnotesize\ensuremath{{}^{\{\arabic{CNOTE}\}}}#1}%
    \end{marginpar}%
    {\color{Orange}\ensuremath{{}^{\{\arabic{CNOTE}\}}}}
    {\color{Green} #2}
  \fi%
}
\makeatother

\usepackage[utf8]{inputenc}
\usepackage[normalem]{ulem}
\newcommand{\rs}[2]{{\color{blue} #2}}

\begin{document}
\bibliographyunit[\chapter]

\title{The strength of the interaction between quarks and gluons
  \protect}
\subtitle{\small{(ALPHA collaboration)}}


\author[1,2]{Mattia~Dalla~Brida}
\author[3]{Roman H{\"o}llwieser}
\author[3]{Francesco Knechtli}
\author[3]{Tomasz Korzec} %
\author[4]{Alberto~Ramos} 
\author[5]{Stefan~Sint}%
\author[6,7]{Rainer~Sommer}
\affil[1]{Dipartimento di Fisica, Universit\`a di Milano-Bicocca, Piazza della Scienza 3, I-20126 Milano, Italy}
\affil[2]{INFN Milano-Bicocca, Piazza della Scienza 3, Milan, I-20126, Italy}
\affil[3]{Department of Physics, Bergische Universit\"at Wuppertal, Gau{\ss}str. 20, 42119 Wuppertal, Germany}%
\affil[4]{Instituto de F\'isica Corpuscular (IFIC), CSIC-Universitat de Valencia, 46071, Valencia, Spain}
\affil[5]{School of Mathematics and Hamilton Mathematics Institute, Trinity College Dublin, Dublin 2, Ireland}%
\affil[6]{John von Neumann Institute for Computing (NIC), DESY, Platanenallee~6, 15738~Zeuthen, Germany}
\affil[7]{Institut~f\"ur~Physik, Humboldt-Universit\"at~zu~Berlin, Newtonstr.~15, 12489~Berlin, Germany}

\abstract{
Modern particle physics experiments, e.g. at the Large Hadron Collider (LHC)
at CERN,  crucially depend on the precise description of the scattering processes
in terms of the known fundamental forces. This is limited by our current understanding
of the strong nuclear force, as quantified by the strong coupling,
$\alpha_s$, between quarks and gluons.  
Relating $\alpha_s$ to experiments poses a major challenge
as the strong interactions lead to the confinement of quarks
and gluons inside hadronic bound states. At high energies,
however, the strong interactions become weaker (``asymptotic freedom’’)
and thus amenable to an expansion in powers of the coupling.
Attempts to relate both regimes usually rely on modeling of
the bound state problem in one way or another.
Using large scale numerical simulations of a first
principles formulation of Quantum Chromodynamics on a space-time lattice,
we have carried out a model-independent determination of $\alpha_s$
with unprecedented precision. The uncertainty, about half that of all other results
combined, originates predominantly from the statistical Monte Carlo
evaluation and has a clear probabilistic interpretation.
The result for $\alpha_s$ describes a variety of physical phenomena over
a wide range of energy scales. If used as input information,
it will enable significantly improved analyses of many high energy experiments,
by removing an important source of theoretical uncertainty.
This will increase the likelihood to uncover small effects of
yet unknown physics, and enable stringent precision tests of the Standard Model.
In summary, this result boosts the discovery potential of the LHC and future colliders,
and the methods developed in this work pave the way for even higher
precision in the future.  
}


%


\maketitle

\addcontentsline{toc}{section}{Main}

\begin{bibunit}

\subsubsection*{Properties of the strong interactions}

At the fundamental level, the strong nuclear force between nucleons
arises from Quantum Chromo Dynamics (QCD), a quantum field theory
formulated in terms of their ``colour-charged'' elementary
constituents, the quarks and gluons.  The strength of the
quark-gluon coupling, $\alpha_{\rm x}(\mu)$, depends on
the energy scale, $\mu$, of the interaction and also on its detailed
definition, summarized as ``scheme'', ${\rm x}$. 

Quarks do not appear as free particles but are
confined in colour-neutral hadronic states such as protons, neutrons
or $\pi$-mesons. Quark confinement arises due to $\alpha_{\rm x}(\mu)$
becoming large at typical hadronic energy scales below a $\text{GeV}$,
where $\mu$ approaches $\Lambda_{\rm x}$, the characteristic scale or
$\Lambda$-parameter of QCD.  The absence of free quarks makes it
impossible to measure the coupling directly in experiment.

\begin{figure} \centering
\includegraphics[width=\textwidth]{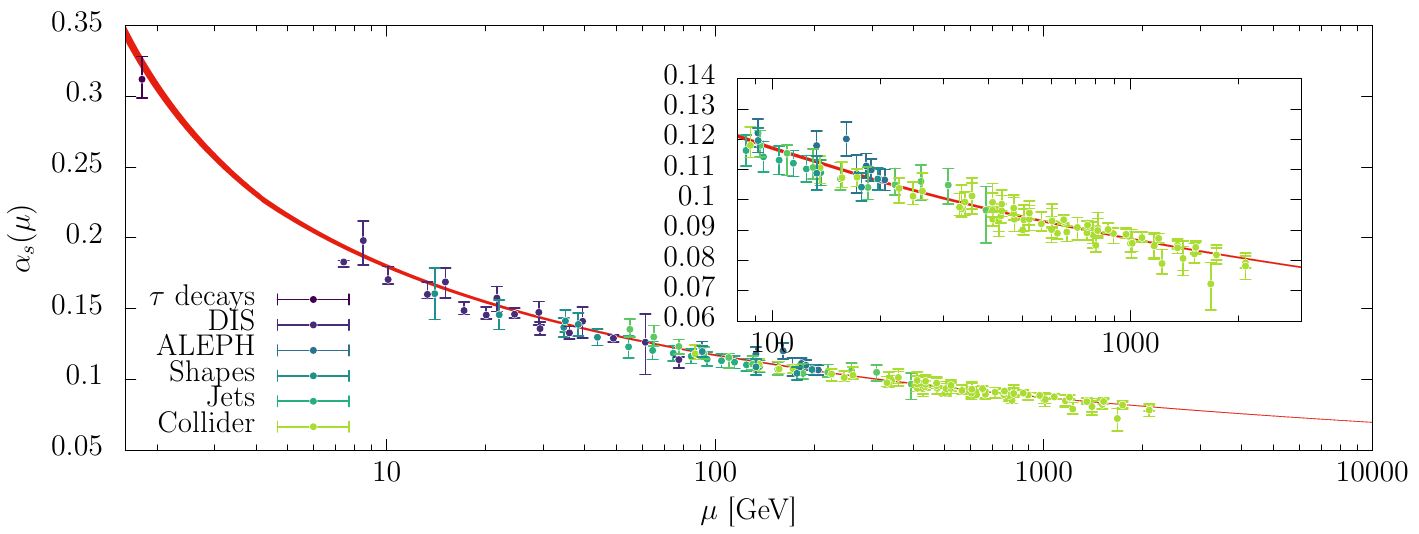}
  \caption{The strong coupling for a wide range of energy scales, as
determined from our result for $\Lambda_{\rm QCD}$, is represented by the red band. Also shown
are experimental determinations from various processes with their
uncertainties as quoted by the Particle Data Group (PDG)~\cite{ParticleDataGroup:2024cfk}.}
  \label{fig:alphas}
\end{figure}

The scale dependence of the strong coupling is described by its
$\beta$-function,
\begin{equation}
  \mu \frac{\rmd}{\rmd\mu} \alpha_{\rm x}(\mu) =
\beta_{\rm x}(\alpha_{\rm x}(\mu))\,.
\label{e:RGeq}
\end{equation} which has an expansion of the form $\beta_{\rm x}(\alpha_{\rm
  x}) = -\beta_0
\alpha_{\rm x}^2 -\beta_1 \alpha_{\rm x}^3 + \rmO(\alpha_{\rm x}^4)$, with leading {\em
  positive} coefficients 
$\beta_0,\beta_1$ which are independent of the scheme. This implies 
that $\alpha_{\rm x}(\mu)$ ``runs'' with the scale $\mu$, decreasing for increasing $\mu$, with a leading
behaviour $\propto 1/\ln(\mu/\Lambda_{\rm x})$ as shown in \fig{fig:alphas}.
This phenomenon, known as asymptotic
freedom~\cite{Gross:1973id,Politzer:1973fx} implies that 
perturbative series expansions in powers of the strong coupling become
accurate at high energies, as exemplified by the $\beta$-function
itself. The scheme independence of the leading coefficients, $\beta_0,\beta_1$,
implies that the asymptotic scale dependence is universal
and $\Lambda$-parameters of different schemes are simply related by
exactly calculable constants. 

Conventionally, $\Lambda_{\rm QCD}\equiv\Lambda_\msbar$
is used for a reference, where $\msbar$ stands for the modified minimal subtraction scheme~\cite{Bardeen:1978yd}. Knowledge of $\Lambda_{\rm QCD}$ and the  
$\beta$-function is equivalent to knowing the coupling at any given scale $\mu$. 
In the $\msbar$ scheme, the expansion coefficients of
$\beta_{\overline{\rm MS}}$ are known to high order including $\beta_4$,
i.e.~5-loop order~\cite{vanRitbergen:1997va,Czakon:2004bu,Baikov:2016tgj,Luthe:2016ima,Herzog:2017ohr}, so that
the scale dependence of $\alpha_{\overline{\rm MS}}(\mu)$ can be 
accurately predicted down to $\mu$ of order 1~GeV. In this paper we 
determine $\Lambda_{\rm QCD}$ using large scale simulations.

With decades of theoretical and experimental efforts to parameterize the effects of
confinement, and to identify observables where these effects are
minimized, the theoretically expected scale dependence could be
verified from experiments as illustrated by the data points in
\fig{fig:alphas}. However, the uncertainties due to the 
required modelling of confinement are hard to quantify and this often means
that increased statistics is of limited use.

\subsubsection*{The r\^ole of lattice QCD}

The modelling of confinement is entirely by-passed in lattice QCD, a
genuinely non-perturbative formulation of QCD on a (Euclidean)
space-time lattice with spacing $a$. Quark and gluon fields are
sampled on the lattice points and edges, respectively.
If the space-time volume is finite, the number of QCD degrees of freedom is
reduced to a finite albeit large number, enabling the numerical
evaluation of observables by large scale computer
\rs{}{``}simulations\rs{}{''}.  Predictions for hadronic observables,
such as the proton's mass $m_p$ or the leptonic decay width of 
$\pi$-mesons, can be obtained for a given choice of the Lagrangian parameters,
the bare quark masses and bare coupling, with the latter controlling the lattice
spacing. In order to make contact with the natural world, one needs to
take the continuum limit, $a\to 0$, based on numerical data for a
range of $a$-values, finite volume effects must be controlled, and the
bare quark masses must be tuned in order to match the physical values
of the chosen experimental input.

In a volume large enough to accommodate hadrons, the typical momentum
cutoff $\pi/a$ is in the range
$6-15~\text{GeV}$. This is one order of magnitude below 
the universal large energy region, where low order perturbation theory is
accurate. Together with the basic requirement that physical scales
have to be well below the cutoff, $\mu \ll \pi/a$, a large volume
approach to determine the strong coupling would require lattices with
significantly more than 100
million lattice points (the current state of the art), along with
orders of magnitude more computational resources than those presently available. 
Consequently we use a scheme for the
running coupling where the energy scale is given by the size of
the simulated world, $\mu=1/L$. Small volumes probe the
high energy regime of QCD, while large volumes probe low energy scales.
The energy dependence of the coupling is obtained by simulating pairs of lattices
with extents $L/a$ and $2L/a$, and a subsequent continuum
extrapolation \cite{Luscher:1991wu}. This relates the values of the coupling
separated by a factor 2 in scale. Iterating this step scaling $n$
times, a scale change of $2^n$ is achieved. For QCD with $\Nf=3$
flavours the method was developed and applied \cite{Fritzsch:2013je,
Brida:2016flw,DallaBrida:2016kgh,Bruno:2017gxd} over many years. Here
we reach a further significant increase in precision and the important
confirmation that the continuum limit is approached smoothly.  
Furthermore we complement this direct approach with the ``decoupling
technique'' of refs.~\cite{DallaBrida:2019mqg,DallaBrida:2022eua} with
essential improvements, as explained below.


\subsubsection*{Scale setting using $\sqrt{t_0}$}

Results from lattice simulations come in units
of the lattice spacing $a$, e.g. $\hat{m}_p=am_p$. In order to express them in physical
(energy) units, the units of the lattice spacing $a$ must be established through 
an experimental input, e.g. $a=\hat{m}_p/m_p^{\rm exp}$.
To minimize the uncertainty from the conversion between lattice and physical
units, we introduce an intermediate step, by first
relating the experimental input to a technical \mbox{(length-)} scale,
$\sqrt{t_0}$, derived from the gradient
flow (GF)~\cite{Luscher:2010iy}.  Together with colleagues in the CLS
consortium, we obtained \cite{Bruno:2016plf,Strassberger:2021tsu}
$\sqrt{t_0} = 0.1443(7)(13)$~fm from the leptonic decay rates of $\pi$
and $K$ mesons.  Nominally more precise values are available from the
literature, which differ by the choice of experimental input or
discretisation of QCD, and some results include the heavier charm
quark in the
simulations~\cite{Alexandrou:2021bfr,Miller:2020evg,Bazavov:2015yea,Dowdall:2013rya,RQCD:2022xux,Blum:2014tka,Borsanyi:2012zs}.
We will use $\sqrt{t_0} =
0.1434(18)$~fm, which includes a rather generous uncertainty of
$1.3\%$ that covers all precise results entering the FLAG
average~\cite{Aoki:2021kgd}, while still producing only a subdominant
source of uncertainty when propagated into our result for
$\Lambda_\text{QCD}$.

Our main computation is further split as
\begin{equation}
\Lambda_{\rm QCD}\sqrt{t_0} =\mu_{\rm dec} \sqrt{t_0}\times
\frac{\Lambda_{\rm QCD}}{\mu_{\rm dec}}\,,
\end{equation}
where $\mu_{\rm dec} \approx 800\,\text{MeV}$ denotes the decoupling scale of Refs.~\cite{DallaBrida:2019mqg,DallaBrida:2022eua}.
Both dimensionless factors can be computed with high precision, however,
the second factor presents a major challenge and dominates the error budget. Therefore,
we computed it using two methods with very different systematics: the direct
approach in $\Nf=3$ and the decoupling method.

\subsubsection*{Direct approach in $\Nf=3$ QCD}

We implemented the step-scaling method using two different renormalization schemes 
for the coupling at low and high energy scales, respectively.  In the region from hadronic
$\mu\approx 200$~MeV to intermediate scales $\mu\approx4$~GeV, our
finite volume scheme is based on the gradient flow~\cite{Fritzsch:2013je,Luscher:2010iy} and is closely related to the
low energy scale $\sqrt{t_0}$. 
Altogether our dataset includes 98 simulations at 10 different 
volumes $L$ in the range $1/L \approx 0.2 - 4\,\GeV$. Compared
to \cite{DallaBrida:2022eua,Brida:2016flw} our new analysis includes
a very fine lattice spacing, with $a/L=1/64$. This allows us to improve
the precision and perform crucial checks on the previous continuum
extrapolation. We
implicitly define an energy scale, $\mu_\text{dec}$, by prescribing the value
$\alpha_\text{GF}(\mu_\text{dec})={3.949}/{(4\pi)}$. We then determine
$\mu_\text{dec}\sqrt{t_0} = 1.647(17)$ which implies $\mu_\text{dec} =
802(13)\,\text{MeV}$. For energies
above $\mu_{\rm dec}$, we combine these results with our previous simulations of
the high-energy regime in the SF scheme~\cite{Bruno:2017gxd}. 
These include more than 40 simulations at 8 values of the volume $L$, 
that allow to explore energy scales $1/L\approx
4 - 140\,\GeV$ non-perturbatively. 
An extensive analysis of the continuum limit together with a detailed
exploration of the asymptotic high energy
regime~\cite{DallaBrida:2018rfy} leads to $\Lambda_{\rm QCD}/\mu_{\rm
  dec} = 0.432(11)$, which translates to our final result for the direct method, 
$\Lambda_\text{QCD}= 347(11)\,\text{MeV}$.
Even though a further error reduction, especially in the high energy part,
appears feasible, we decided to develop an alternative: the decoupling method. 
It is computationally more efficient and, even more importantly, affected by
very different systematics, thereby providing a strong cross-check.

\begin{figure} \centering
	\includegraphics[width=0.49\textwidth]{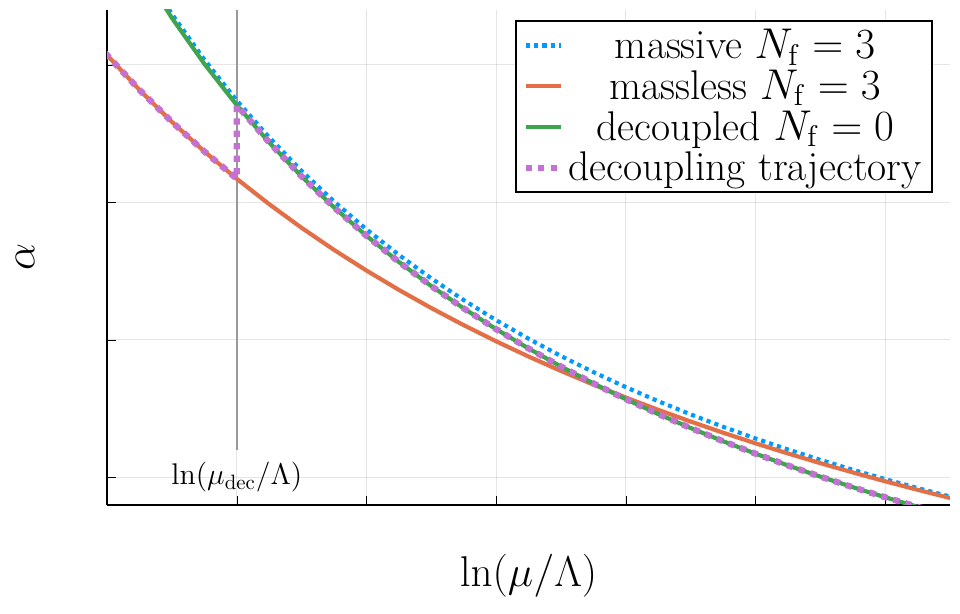}
\hfill \includegraphics[width=0.49\textwidth]{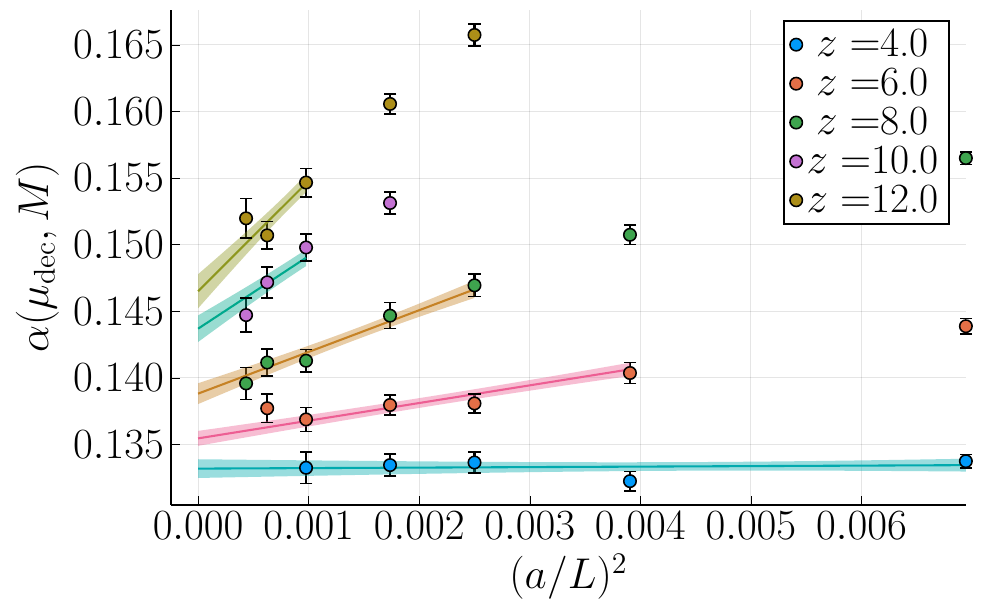}
\caption{\textbf{Left: } Ilustration of the decoupling of three heavy
  quarks with large mass as described in the model of
  section~\ref{sec:1loopMassiveGFcoupling}. For energies $\mu \ll M$
  the massive coupling runs 
	like the pure gauge coupling, while for $\mu \gg M$ the coupling runs
	like the massless three flavor coupling.
	\textbf{Right:} Continuum extrapolation of the massive
	coupling $\alpha(\mu, M)$ for $z = M/\mu_{\rm dec} =
	4,6,8,10,12$.  Even with the conservative cutoff in the data,
	$(aM)^2 < 0.16$, the extrapolated continuum values are still very
	precise.}
\label{fig:cont}
\end{figure}

\subsubsection*{The decoupling method}

The idea is based on the
following observation~\cite{DallaBrida:2019mqg}.  If one increases the
masses of the quarks in a Gedanken-experiment, eventually the
low-lying spectrum of QCD matches the spectrum of the pure gauge
theory, where quarks are absent; we say they are decoupled. In this
way, QCD is connected with the pure gauge theory, the theory without
any quarks.  Since the latter is easy to simulate, better precision
can be achieved compared with QCD~\cite{DallaBrida:2019wur}.
The exact connection requires the fundamental scale of 
the pure gauge theory, $\Lambda^{(0)}$ to be adjusted appropriately,
$\Lambda_{\msbar}^{(0)}=P(M/\Lambda_{\msbar}^{(3)})\,
\Lambda_{\msbar}^{(3)}$, where in $\Lambda_{\msbar}^{(\nf)}$ the number of quarks, $\nf$, is indicated. The matching factor $P$ is known
perturbatively to four loop order~\cite{Athenodorou:2018wpk} and
is routinely being used 
to relate $\Lambda^{(3)}_\msbar\to\Lambda^{(4)}_\msbar\to\Lambda^{(5)}_\msbar$,
across the charm and bottom quark thresholds~\cite{Herren:2017osy}. 

Decoupling works up to O$(1/M^2)$ corrections, where
$M$ denotes the RGI quark mass (see methods). Detailed studies have
shown~\cite{Athenodorou:2018wpk} that the corrections are small already at masses
of the order of the charm quark mass ($M_{\rm c}\approx 1.4$~GeV).
Here we use masses in the range $M \sim 3-10$~GeV, allowing us to explore
the approach $M\to \infty$  in detail and safely match QCD with the
pure gauge theory.  Again we work in the GF scheme but in
QCD with three degenerate heavy quarks. 

To illustrate the procedure,
the running of this massive coupling $\alpha_{\rm GF}(\mu,M)$ is shown
schematically in \fig{fig:cont} (left). For energies well below their
mass, the heavy quarks are decoupled and $\alpha_{\rm GF}$ runs like in pure
gauge theory (green line) while at $\mu$ far above their mass the running is
governed by the massless $\beta$-function; it is slowed down.  

We now follow the magenta trajectory in the figure. Below $\mu_{\rm dec}$  we 
reuse the running in the mass-less theory.  Then, at
fixed $\mu=\mu_{\rm dec}$, we increase the mass of all three quarks
artificially to very high values, eventually to $M\approx10$~GeV,
following the vertical part of the magenta trajectory. At the resulting
value of the massive coupling, we switch to the pure gauge theory and
run to large $\mu$ where $\Lambda^{(0)}_\mathrm{GF}$ is obtained.
Converted (exactly) to the $\msbar$ scheme, we then use the accurate
high order relation $P(M/\Lambda_\msbar^{(3)})$ between the $\Lambda$-parameters with and without
quarks to revert to $\Lambda_\msbar^{(3)}=\Lambda_\text{QCD}$. 

The main
challenge is the continuum extrapolation of the massive coupling from
our simulation results at finite $a$.  On the one hand the quark mass
has to be large for the decoupling approximation to be as accurate as
possible.  On the other hand the mass has to be below the momentum
cutoff $\sim \pi/a$ of the lattice. In other words we need $aM \ll 1$
and a good understanding of the asymptotic behavior of discretisation
effects close to the continuum limit.  To this end, we determined an
improved discretisation such that linear effects in the lattice
spacing are absent.  With the result of \cite{DallaBrida:2023yka} we
could cancel the dangerous terms proportional to $aM$.  Next,
sufficiently small lattice spacings were simulated and we performed
a
combined extrapolation of the results for different quark masses and
different lattice spacings. Analyzing the theory for discretization
effects \cite{Husung:2019ytz} in an expansion in $1/M$, one arrives at an
asymptotic form of the discretization errors with only two free
parameters. This form fits the data, \fig{fig:cont}, remarkably well.
The fit tells us that the coupling changes from
$\alpha_\mathrm{GF}(\mu,M) =       0.4184(22)$ for $M\approx 3$ GeV
to $\alpha_\mathrm{GF}(\mu, M) =       0.4600(41)$ for $M \approx 10$ GeV in
continuum QCD.

We then matched to the pure gauge theory by equating $\alpha(\mu;
M)$ with the $\Nf=0$ coupling. Using previous results in the pure
gauge theory from~\cite{DallaBrida:2019wur}, we arrive at an estimate
of $\Lambda_{\rm QCD}/\mu_{\rm dec}$ at each value of the quark mass.  These
estimates should all agree up to the mentioned O$(1/M^2)$ corrections.
Indeed the numbers are very close: for quark masses between 5 to 10
GeV, $\Lambda_{\rm QCD}/\mu_{\rm dec}$ varies by 5\%. They also follow the 
expected $c_0+c_1 M^{-2}$ behavior.  An extrapolation with this form
thus yields our final number $\Lambda_{\rm QCD}/\mu_{\rm dec} =
0.4264(97)$ in the three flavor theory from the decoupling strategy. 
Together with the value for $\mu_{\rm dec}$, we get
$\Lambda_\text{QCD} = 341.9(9.6)$ MeV.  The 
uncertainty covers the statistical errors and several variations of
the functional form employed in the continuum, $a\to0$, and
decoupling, $M\to \infty$, extrapolations.  It also includes the
uncertainty of the conversion from $\Lambda_\msbar$ of the pure gauge
theory to $\Lambda_\text{QCD}$.

\subsubsection*{Final result and concluding remarks}

\begin{figure} \centering
\includegraphics[width=\textwidth]{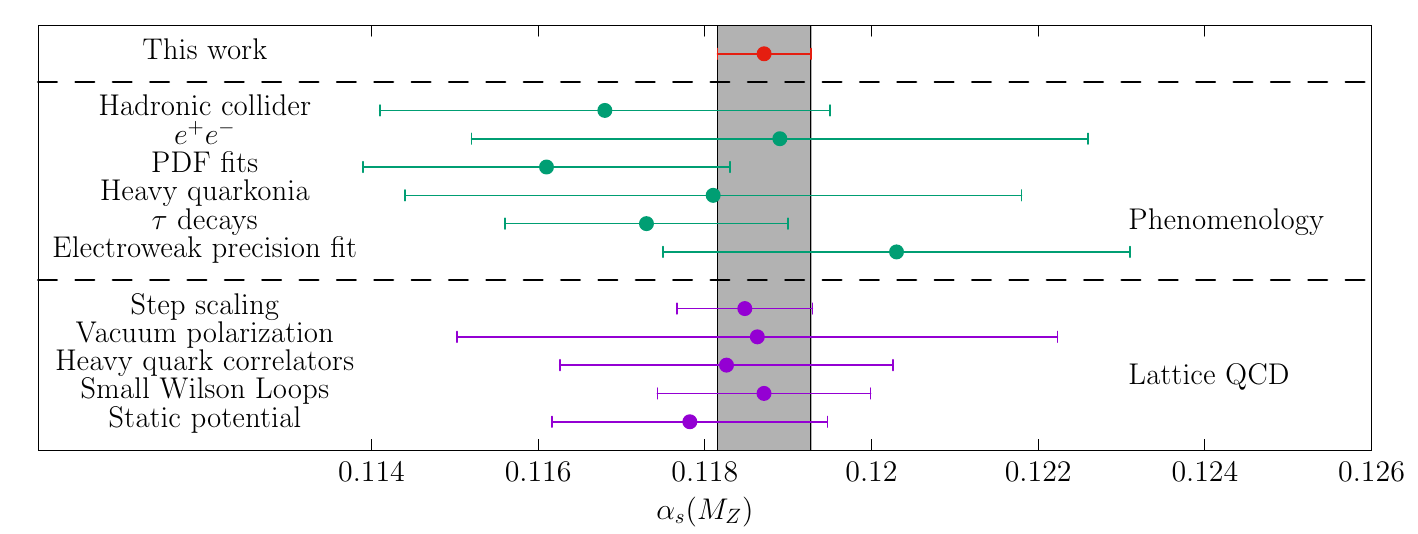}
  \caption{Our result for $\alpha_s(m_Z)$ compared with other results
from the literature (cf.~text).}
  \label{fig:comparison}
\end{figure}

Both the above
methods to extract $\Lambda$ have uncertainties dominated by
statistics.  Theoretical uncertainties, in particular those related
to the use of perturbation theory are subdominant. The systematics
are also very different in both methods, and their agreement further
corroborates the robustness of our methodology. An average is
justified and leads to
\begin{equation}
  \label{eq:LambdaNf3Final}	
  \Lambda_{\rm QCD} = 343.9(8.4)\, {\rm MeV}\,.
\end{equation}
We then include the charm and bottom contributions in
the running in the $\msbar$ scheme in perturbation theory and arrive
at the curve in figure~\ref{fig:alphas}. In particular we have,
\begin{equation}
  \alpha_s(m_Z) = 0.11873(56)\,, 
\end{equation}
where $\alpha_s = \alpha_\msbar^{(\Nf=5)}$.
Figure~\ref{fig:comparison} shows our result compared with numbers
from other strategies. Most of them have uncertainties dominated by
theoretical or systematic effects, as quoted by the PDG for the
phenomenology results and by FLAG for lattice results~\cite{FlavourLatticeAveragingGroupFLAG:2024oxs}.  
An exception is the category labeled ``Step Scaling'', that uses the
methods developed along the years by the ALPHA collaboration.  This
result is dominated by our earlier computation~\cite{Bruno:2017gxd}.

What have we learned?  Recall that QCD is a complicated non-linear
theory with the observed particles completely different from the
fundamental quanta in the Lagrangian.  Still, surprisingly, we are
able to determine the intrinsic scale $\Lambda_\text{QCD}$ of the
theory and equivalently the coupling between quarks and gluons. A
conceptual achievement beyond the mere precision of $\alpha_s$ is that
it is determined with experimental low energy input from the decays
and masses of hadrons which are all bound states of quarks and
gluons. Fig.~1 compares the resulting coupling with phenomenological
determinations.  While the latter have their issues, the overall
qualitative agreement confirms QCD as the single theory of the strong
interactions at all energies both small and large compared to
$\Lambda_\text{QCD}$.  Thus there is very little room for any
modifications/additions to the theory of the strong interactions.  The
size of the error band demonstrates that our value can be used to
better understand the physics involved in these processes, and to
boost the discovery potential of the LHC. For example, our sub-percent precision in $\alpha_s$ is needed to control predictions such as Higgs production through gluon-fusion at the high luminosity LHC~\cite{doi:10.1142/9789813238053_0007} and 
as a constraint in the determination of the hadronic Parton
Distribution Functions (PDFs)~\cite{ParticleDataGroup:2024cfk,CMS:2019oeb}.


\putbib  
\end{bibunit}

\begin{bibunit}

\section{Methods}
\label{sec:methods}
\subsection{The scale of QCD and the $\Lambda$ parameter}
\label{sec:LambdaParam}

We consider QCD with $\Nf$ quark flavours and show how an intrinsic
scale can be defined, the QCD $\Lambda$-parameter. In what follows we adopt the conventions 
of our past papers and use the coupling $\bar g_{\rm x}(\mu) = \sqrt{4\pi\, \alpha_{\rm x}(\mu)}$. 
The $\beta$-function, as a function of $\bar{g}_{\rm x}$, is given by,
\begin{equation}
	\label{eq:beta-function}
   \mu\frac{\partial\bar{g}_{\rm x}(\mu)}{\partial\mu} = \beta_{\rm x}(\bar{g}_{\rm x}),
\end{equation}
We emphasize that $\beta_{\rm x}$ is non-perturbatively defined if this is the case for the coupling $\bar{g}_{\rm x}$.
As mentioned earlier, the first two coefficients in the expansion $\beta_{\rm x}(g) = - b_0 g^3 -b_1 g^5 +\rmO(g^7)$,
\begin{equation}
  b_0= \left(11-\frac23\Nf\right)\times(4\pi)^{-2},\quad b_1 = \left(102-\frac{38}{3}\Nf\right)\times (4\pi)^{-4},
\end{equation}
are renormalization-scheme independent ($\beta_k$ of Eq.~(\ref{e:RGeq}) is proportional to $b_k$, for $k=0,1,\ldots$).
This asymptotic behaviour for small couplings allows us to integrate the differential equation (\ref{eq:beta-function})
between the scales $\mu$ and $\mu'$ as follows,
\begin{equation}
  \dfrac{\mu'}{\mu} = \exp\int\limits_{\bar{g}_{\rm x}(\mu)}^{\bar{g}_{\rm x}(\mu')} \dfrac{\rmd g}{\beta_{\rm x}(g)} = 
  \dfrac{\varphi_{\rm x}(\bar{g}_{\rm x}(\mu))}{\varphi_{\rm x}(\bar{g}_{\rm x}(\mu'))}\, .
  \label{eq:scale-ratio}
\end{equation}
Here, the function $\varphi_{\rm x}$ is given by,
\begin{equation}
    \varphi_{\rm x}^{}(\gbar_{\rm x}) = ( b_0 \gbar_{\rm x}^2 )^{-b_1/(2b_0^2)} 
        \rme^{-1/(2b_0 \gbar_{\rm x}^2)} \times \exp\left\{-\int\limits_0^{\gbar_{\rm x}} \rmd g\ 
        \left[\frac{1}{\beta_{\rm x}(g)} 
             +\frac{1}{b_0g^3} - \frac{b_1}{b_0^2g} \right] \right\}\,.
  \label{eq:varphi}           
\end{equation}
Note that, by design, the integrand in the exponent is regular at zero coupling. 
Furthermore, the combination $\mu\varphi_{\rm x}(\bar{g}_{\rm x}(\mu))$
is independent of $\mu$, has units of energy and is known as the $\Lambda$-parameter of QCD,
\begin{equation}
  \Lambda_{\rm x}^{(\Nf)} = \mu\varphi_{\rm x}(\bar{g}_{\rm x}(\mu)) = \mu' \varphi_{\rm x}(\bar{g}_{\rm x}(\mu')). \label{e:Lambda_from_mu}
\end{equation}
Since these equations are exact, the $\Lambda$-parameter can be evaluated at any scale $\mu$,
provided the integral in the exponent of $\varphi_{\rm x}$ can be evaluated reliably. 
This poses a particular challenge for lattice QCD: the experimental observables used to
fix the free parameters, i.e. the quark masses and the lattice spacing, are measured at very low energies, where
QCD is clearly non-perturbative. Matching to hadronic physics thus requires a hadronic low energy scale, $\mu_\text{had}$. 
However, the  integral extends to zero coupling, i.e.~an infinitely large energy scale.
The idea then is to split the computation into two factors,  
\begin{equation}
  \dfrac{\Lambda_{\rm x}^{(\Nf)}}{\mu_\text{had}}  = \dfrac{\mu_\text{pert}}{\mu_\text{had}} \times\varphi_{\rm x}(\bar{g}_{\rm x}(\mu_\text{pert}))\,.
   \label{e:LambdaRatio}
\end{equation}
where $\mu_\text{pert} \gg \mu_\text{had}$ is a large scale, in the perturbative regime of QCD.
One then uses the non-perturbative step-scaling approach (s.~below) to
determine the large scale ratio (first factor), together with the values of the
coupling at both scales. 
The second factor is evaluated in perturbation theory, by inserting
the perturbative expansion  of the $\beta$-function, which is, in the $\msbar$-scheme, known up to 5-loop order.
Here we use the SF scheme~\cite{Luscher:1992an} to define $\mu_\text{pert}$ through 
$\bar{g}^2_{\rm SF}(\mu_{\rm pert}) = 1.085(4)$, which yields the value
\begin{equation}
  \label{eq:lamsf} 
  \frac{\Lambda^{(3)}_{\rm SF}}{\mu_{\rm pert}} = 0.003788(88)\, .
\end{equation}
While the central value can be easily calculated from Eq.~(\ref{eq:varphi}), given the value of the 
coupling and the 3-loop approximation of the SF $\beta$-function, the error estimate
requires a careful comparison of perturbative and non-perturbative scale evolution, where
the latter is obtained from the step-scaling procedure (s. below). 
In physical units, the perturbative matching scale $\mu_\text{pert}$ is found to be approximately $70$ GeV.
By varying this scale in the range  $4 - 128$ GeV~\cite{DallaBrida:2018rfy}, 
we performed extensive tests of perturbation theory and our final choice combines moderate statistical errors 
with significantly smaller systematic errors.

The $\Lambda$-parameter is scheme dependent. 
Given two mass-independent schemes ${\rm x}$ and ${\rm y}$, with the respective
couplings related perturbatively by 
\begin{equation}
    g_{\rm x}^2 = g_{{\rm y}}^2 + c_{{\rm x}{\rm y}}g_{{\rm y}}^4 + \ldots 
  \end{equation}
  we have the exact relation
  \begin{equation}
\Lambda_{{\rm x}}/\Lambda_{\rm y} = \exp(c_{{\rm x}{\rm y}}/2b_0), \label{e:LambdaRatio}
  \end{equation}
i.e.  the scheme-dependence is
completely encoded in the  
perturbative one-loop coefficients relating the respective
couplings. This provides an indirect non-perturbative 
meaning for $\Lambda$-parameters defined in purely perturbative
schemes like the $\msbar$-scheme, 
and $\Lambda_\msbar^{(\Nf)}$ is thus taken as the reference scale in
QCD. For our particular case, we have $\Lambda^{(3)}_{\rm
  SF}/\Lambda^{(3)}_{\overline{\rm MS} } = 0.38286(2)$, which results in
\begin{equation}
  \frac{\Lambda^{(3)}_{\overline{\rm MS} }}{\mu_{\rm pert}} = 0.00989(23)\,.
\end{equation}

\subsection{Finite volume renormalization schemes and step scaling}
\label{sec:StepScaling}

In order to relate physical observables defined at very different scales 
by lattice simulations, the key idea is to define an intermediate 
renormalization scheme in a finite space-time volume, where the linear extent of
the volume, $L$, is used to set the renormalization scale, i.e.~$\mu=1/L$.
Given the renormalized coupling in such a finite volume scheme $s$, one then
proceeds with the computation of the so called step-scaling function (SSF),
\begin{equation}
  \sigma_{\rm x}(u) = \bar g_{\rm x}^2(\mu/2)\Big|_{\bar g_{\rm x}^2(\mu) = u}\,,
\end{equation}
which determines the coupling at scale $\mu/2$ as a function of the coupling at scale $\mu$.
The step-scaling function is closely related to the $\beta$-function,
\begin{equation}
  \label{eq:sigbeta}
   \int\limits^{\sqrt{u}}_{\sqrt{\sigma_{\rm x}(u)}} \dfrac{\rmd g}{\beta_{\rm x}(g)} = \ln 2
\end{equation}
Given a finite volume coupling $\bar{g}_{\rm x}(\mu)$ and its SSF $\sigma_{\rm x}(u)$ for
a range of $u$-values, one obtains the non-perturbative
scale evolution of the coupling towards lower energy scales by setting  $u_0= u_\text{min}$ 
and then iterating
\begin{equation}
  u_{k} = \sigma_{\rm x}(u_{k-1}),\qquad k=1,2,3,\ldots
\end{equation}
until one exits the range of available $u$-values. After 10 steps
one has covered a scale factor of $2^{10}$, i.e.~3 orders of magnitude.
Yet the non-perturbative construction of the SSF can be carried out
without ever dealing with large scale differences. All that is
required are pairs of lattices of linear dimension $L/a$ and $2L/a$, where
the measured value of the coupling on the smaller lattice defines the argument $u$ and
the measured value (at the same bare parameters) on the $2L/a$-lattice defines a lattice 
approximant, $\Sigma_{\rm x}(u,a/L)$ to $\sigma_{\rm x}(u)$. Repeating such 
computations for a range of lattice resolutions, $a/L$, then allows
one to take the continuum limit  
\begin{equation}
    \sigma_{\rm x}(u) = \lim_{a/L\rightarrow 0} \Sigma_{\rm x}(u,a/L)\,. 
\end{equation}
The procedure is repeated for different values $u$ by adjusting the
bare coupling $g_0^2$.   
Note that eq.~(\ref{eq:sigbeta}) can be used to obtain the $\beta$ function. 
First one finds a suitable parametrization of the
$\beta$-function. Several different choices have been made in the 
literature, from simple polynomials, inspired by perturbation
theory~\cite{Brida:2016flw}, to Pad{\'e} ans\"atze, suitable for a
description of data at low and intermediate energy
scales~\cite{DallaBrida:2016kgh}. Once the functional form $\beta_{\rm x}(x)$ is fixed,
the set of parameters are determined using a standard $\chi^2$ fit
\begin{equation}
  \chi^2 = \sum_{i=1}^{N_{\rm pt}} \left( \frac{\ln 2 - \int^{\sqrt{u}}_{\sqrt{\sigma_{\rm x}(u)}} \dfrac{dg}{\beta_{\rm x}(g)}}{\sqrt{[\beta(\sqrt{u})]^{-2}[\delta u]^2 + [\beta(\sqrt{\sigma_{\rm x}(u)})]^{-2} [\delta\sigma_{\rm x}(u)]^2}} \right)^2\, .
\end{equation}

Knowledge of $\beta_{\rm x}(g)$ then allows us to compute the ratio of the scales
associated with two values of the coupling,
\begin{equation}
  \ln \frac{\mu_1}{\mu_2} = \int_{\bar g_{\rm x}(\mu_1)}^{\bar g_{\rm x}(\mu_2)} \frac{{\rm d} x}{\beta_{\rm x}(x)}\,.  
\end{equation}
While the step-scaling strategy is straightforward in principle, 
its practicality depends on a  number of technical details which have been developed over the
last 30 years. Perturbation theory in finite volume is non-standard and can be
very intricate, depending on the chosen boundary conditions. Using Schr\"odinger
Functional (SF) boundary conditions and the standard SF coupling~\cite{Sint:1993un,Luscher:1992an}, 
the perturbative conversion to the $\msbar$-scheme has been pushed to 2-loop order, 
so that the 3-loop $\beta$-function is known~\cite{Bode:1999sm}.
However, the standard SF coupling is not so well-suited at low energies, 
due to the increase of its variance for larger physical volumes. 
Schemes based on the gradient flow (GF) are much better adapted to
this regime. A strategy using both couplings and a non-perturbative matching between them at
an intermediate scale has been devised for QCD with $\Nf=3$ quark flavours in ref.~\cite{Bruno:2017gxd}. 
We refer to this reference for the precise definition of the couplings in the SF and GF schemes.

These practical issues lead to a further splitting of the ratio $\mu_{\rm pert} /
\mu_{\rm had}$ of Eq.~(\ref{e:LambdaRatio}) into two
factors,
\begin{equation}
  \frac{\mu_{\rm pert} }{\mu_{\rm had}} = \frac{\mu_{0} }{\mu_{\rm had}} \times
  \frac{\mu_{\rm pert} }{\mu_{0}}\,, 
\end{equation}
where $\mu_0$ denotes the scale at which the GF and SF couplings are matched non-perturbatively.
The first factor (i.e.~the low energy part), is computed using the
Gradient Flow scheme with Schr\"odinger Functional boundary
conditions~\cite{Fritzsch:2013je}, while the second factor is computed
using the SF scheme~\cite{Luscher:1991wu,Sint:1993un}. 
This strategy combines the advantages of both schemes: the high statistical precision
of the GF scheme at low scales and the available analytic control in perturbation theory 
for the SF scheme. 

The second factor comes from the data used in~\cite{Brida:2016flw}
together with the detailed study of the linear scaling violations and
the matching with perturbation theory described in~\cite{DallaBrida:2018rfy}. 
Defining $\mu_0$ implicitely through  $\bar g^2_{\rm SF}(\mu_0) =
2.012$, we iterate the step scaling function 4 times,
to arrive at the scale $\mu_{\rm pert}\approx
70$ GeV, and, therefore, we have,
\begin{equation}
    \frac{\mu_{\rm pert}}{\mu_{0}}  = 16\,.
\end{equation}

At low energies, the simulations of~\cite{DallaBrida:2016kgh} have been significantly
improved with simulations close to the continuum limit (see table~\ref{tab:ssf_data}). 
Precise values in the continuum limit can be obtained for the step
scaling function $\sigma(u)$, and these can be used to obtain the
$\beta$-function. 
Defining the scales $\bar g^2_{\rm GF}(\mu_{\rm had}) = 11.31$ and
$\bar g^2_{\rm GF}(\mu_{0}/2) = 2.6723(64)$, we get the ratio,
\begin{equation}
    \frac{\mu_{0}}{\mu_{\rm had}}  =   21.86(33)\,.
\end{equation}
so that $\mu_\text{pert}/\mu_\text{had} = 16\times 21.86(33) = 349.8(5.3)$.

What remains to be done is the matching of $\mu_\text{had}$ to the experimental input.
The scale $\mu_\text{had}$ refers to massless QCD, as the
coupling is renormalized in the chiral limit. To obtain $\mu_\text{had}$ in MeV,
we pass via the technical scale $\sqrt{t_0}$. Using an average of different works (including our own) we obtain
$\sqrt{t_0} = 0.1434(18)$ fm, determined using different experimental inputs 
(mainly meson decay constants and hadron masses). 
The rather generous error of $1.3\%$ covers all the differences, while still be sufficiently small to 
remain subdominant in the total error for the $\Lambda$-parameter.

Relating $\mu_{\rm had}$ to the technical scale $\sqrt{t_0}$ is then done in two steps,
\begin{equation}
  \mu_{\rm had} = \frac{1}{\sqrt{t_0}}\times \sqrt{\frac{t_0}{t_0^\star}} \times
  \left( \sqrt{t_0^\star}\times  \mu_{\rm had}  \right) = 
  200.5(3.0)\, {\rm MeV}\,,
\end{equation}
where $t_0^\star$ is defined as the value of $t_0$ in QCD in the flavour symmetric limit,
with Pions and Kaons of equal masses of about $410\,\MeV$.

Here we have used~\cite{RQCD:2022xux,Strassberger:2021tsu} that $\sqrt{t_0/t_0^\star} =
1.0003(30)$, and, from~\cite{Bruno:2017gxd}, $\sqrt{t_0^\star}\times  \mu_{\rm had} = 0.146(11)$.

To summarize, the step scaling procedure has allowed us to connect
the scales $\mu_{\rm had}\approx 200$ MeV with $\mu_{\rm pert}\approx
70,000$ MeV, thereby bridging a scale factor of 350 without relying on perturbation theory.

\subsection{Decoupling of heavy quarks}

\newcommand{\mudec}{\mu_\mathrm{dec}}
\newcommand{\GF}{\mathrm{GF}}
\newcommand{\Lambdal}{\Lambda^{(0)}}
\newcommand{\eff}{\mathrm{eff}}

The basic physics principles of the decoupling strategy are
summarized in the equation
\begin{equation}
  \label{eq:rhoeq}
  \rho
  P \left( z/\rho\right) =
  \frac{\Lambda^{(0)}_{\overline{\rm MS} }}{\mu_{\rm dec}}
  =
  \frac{\Lambda^{(0)}_{\overline{\rm MS}}}{\Lambda^{(0)}_\text{GF}} \times \varphi_{\rm GF}^{(0)}\left(\gbar_\mathrm{GF}(\mudec,M)\right)\,. 
\end{equation}
In the following we explain the various elements one by one. First, the function $P$ 
is the ratio of $\Lambda^{(0)}$ and $\Lambda^{(3)}$ in
the $\msbar$ scheme, where it can be extracted from the
4-loop matching of the coupling and five-loop running 
with very high accuracy. Its argument is the ratio $M/\Lambda_{\msbar}^{(3)}=z/\rho$, where 
$z=M/\mudec$ involves the renormalization group invariant quark mass,
$M$. We vary $z$ from about 4 to 12 and for each $z$ we extract 
\begin{equation}
\rho = \Lambda^{(3)}_{\overline{\rm MS},\mathrm{eff} }(z)/\mu_{\rm dec}=\Lambda^{(3)}_{\overline{\rm MS} }/\mu_{\rm dec}+\rmO(1/z^2)\,
\end{equation}
as the numerical solution of equation~(\ref{eq:rhoeq}). Combined
with the already known $\mu_{\rm dec}$ we obtain estimates for 
$\Lambda^{(3)}_{\overline{\rm MS},\mathrm{eff} }$.
The function $\varphi_{\rm GF}^{(0)}(g)$ is known rather 
precisely from the step scaling computation \cite{DallaBrida:2019wur},
typically with a 1\% accuracy. It determines the curve labeled ``decoupled'' in fig.~\ref{fig:cont}. The
factor $ {\Lambda^{(0)}_{\overline{\rm
      MS}}}/{\Lambda^{(0)}_\text{GF}}=0.4981(17)$ is known exactly~\cite{DallaBrida:2017tru},
cf.~eq.(\ref{e:LambdaRatio}).  

The values for the massive coupling, $\bar g(\mudec,M)$, where we evaluate $\varphi_{\rm GF}^{(0)}$ are determined by a careful continuum extrapolation. First, as discussed before, it is crucial to remove linear
effects $\sim aM$~\cite{DallaBrida:2023yka}. Second, we used effective field theory (EFT) to derive the leading terms of the expansion of the lattice discretised observable, namely $\gbar_\mathrm{GF}$, in 
$a$ and $1/M$ in the region $aM\ll1$ and $z\gg1$.
The EFTs are Symanzik effective theory as well as the decoupling
expansion, i.e. the pure gauge theory with perturbations by dimension
six operators. The latter provide corrections to the infinite $M$
limit. These EFTs have to be applied in the correct order: the
decoupling expansion is applied to the Symanzik effective theory.    
The crucial information obtained from the EFT analysis is that 
for sufficiently small $a\mudec,aM$ and large $z$, results at finite
lattice spacing are described by  
\begin{equation}
	\bar g_\GF^2(\mudec,M,a) = \bar g_\GF^2(\mudec,M,0)+ 
    p_1 [\alpha_s(a^{-1})]^{\hat \Gamma_{\rm eff}} {a^2\mu_{\rm dec}^2} +
    p_2 [\alpha_s(a^{-1})]^{\hat \Gamma'_{\rm eff}} (aM)^2 \,,
	\label{eq:globalfit}
  \end{equation}
The $a^2$ terms are accompanied by $\alpha_s^{\Gamma}(1/a)$ terms which vary 
logarithmically with the argument. However, the leading powers
$\Gamma\,,\; \Gamma'$ are small~\cite{Husung:2019ytz}. Our explicit tests
with the available data
show that including them or neglecting them has a minor effect
and such variations of the analysis are part of our error budget.

Numerically, the form \eq{eq:globalfit} describes the data very well
for a quite large range of $(a\mudec)^2$ and all $z\geq 4$. In order
to obtain our continuum 
results  $\bar g_\GF^2(\mudec,M,0)$ we impose a  
cut of $(a M)^2<0.16$ but the curves in the figure do extend to larger
values and show that the exact position of the cut is not
relevant. Indeed, the numerical analysis also shows no evidence of
higher powers in $a$ or the mentioned logarithmic corrections. In
summary, the continuum values are solid.  

All that is left is to take the limit $z\to\infty$ of the estimates
$\Lambda_{\overline{\rm MS}, {\rm eff} }^{(3)}(z)/\mu_{\rm dec}$. 
The extrapolation turns our to be smooth, with the data with $z\ge 8$
being compatible with the $M=\infty$ value. 
All data with $z\ge 4$ is well described by a linear functional form
in $1/z^2$. 
Logarithmic corrections to this linear functional form have a very
small effect in the final determination $\Lambda_{\overline{\rm
    MS}}^{(3)}/\mu_{\rm dec}$ (see figure~\ref{fig:lam_variations}).  

We note that our definition of $\gbar_\mathrm{GF}$ takes place in a
finite volume with boundaries in time (as opposed to periodic). In
principle this feature allows for linear corrections in $a$ and
$1/M$. The associated boundary terms in the EFTs are rather simple and
we computed their effects. In our setup with $T=2L$ they are well
below the statistical uncertainties of $\bar g^2_{\rm GF}(\mu_{\rm dec};M)$. 
Note that the uncertainty in the massive coupling contributes a small
amount to the total uncertainty in $\alpha_s$.
 


\clearpage

\section{Extended data figures and tables}
\label{sec:extend-data-figur}

\begin{table}[h!]
  \centering
  \begin{tabular}{ccclccc}
\toprule
  $L/a$ & $u$ & $\Sigma(u, a/L)$ & &$L/a$ & $u$ & $\Sigma(u, a/L)$\\
  \cmidrule(lr){1-3}  \cmidrule(lr){5-7}
8 & 6.5485(61) & 11.452(79) &  & 16 & 6.549(15) & 13.36(14)\\[0pt]
8 & 5.8671(35) & 9.250(66) &  & 16 & 6.037(14) & 11.34(10)\\[0pt]
\textbf{8} & \textbf{5.8648(65)} & \textbf{9.377(25)} &  & 16 & 5.867(14) & 10.91(12)\\[0pt]
8 & 5.3010(35) & 7.953(44) &  & \textbf{16} & \textbf{5.865(11)} & \textbf{10.810(69)}\\[0pt]
8 & 4.4848(24) & 6.207(23) &  & 16 & 5.301(13) & 9.077(75)\\[0pt]
8 & 3.8636(22) & 5.070(16) &  & 16 & 4.4901(77) & 6.868(40)\\[0pt]
8 & 3.2040(20) & 3.968(11) &  & \textbf{16} & \textbf{3.9475(58)} & \textbf{5.689(14)}\\[0pt]
8 & 2.7363(14) & 3.2650(79) &  & 16 & 3.8643(63) & 5.485(22)\\[0pt]
8 & 2.3898(16) & 2.7722(63) &  & 16 & 3.2029(52) & 4.263(16)\\[0pt]
8 & 2.1275(16) & 2.4226(49) &  & 16 & 2.7359(35) & 3.485(11)\\[0pt]
 &  &  &  & 16 & 2.3900(30) & 2.9348(71)\\[0pt]
 &  &  &  & 16 & 2.1257(25) & 2.5360(66)\\[0pt]
12 & 6.5446(81) & 12.87(17) &  &  &  & \\[0pt]
12 & 6.1291(56) & 11.79(13) &  &  &  & \\[0pt]
12 & 5.8730(46) & 10.497(78) &  & \textbf{20} & \textbf{5.857(11)} & \textbf{10.965(77)}\\[0pt]
\textbf{12} & \textbf{5.8697(87)} & \textbf{10.485(36)} &  & \textbf{20} & \textbf{3.9493(63)} & \textbf{5.755(22)}\\[0pt]
12 & 5.2991(36) & 8.686(49) &  &  &  & \\[0pt]
12 & 4.4908(31) & 6.785(36) &  &  &  & \\[0pt]
\textbf{12} & \textbf{3.9461(43)} & \textbf{5.592(11)} &  & \textbf{24} & \textbf{5.877(14)} & \textbf{11.09(11)}\\[0pt]
12 & 3.8666(24) & 5.380(25) &  & \textbf{24} & \textbf{3.9492(62)} & \textbf{5.770(25)}\\[0pt]
12 & 3.2058(17) & 4.180(14) &  &  &  & \\[0pt]
12 & 2.7380(15) & 3.403(11) &  &  &  & \\[0pt]
12 & 2.3903(11) & 2.8963(87) &  & \textbf{32} & \textbf{3.9490(97)} & \textbf{5.844(32)}\\[0pt]
12 & 2.1235(13) & 2.5043(76) &  &  &  & \\[0pt]
\bottomrule 
\end{tabular}


  \caption{Data used to determine the $\beta$-function in $N_{\rm f} =
    3$ QCD. 
    The column labeled $u$ represent the value of the coupling $\bar
    g^2_{\rm GF}(\mu,a/L)$ at a scale $\mu = 1/L$ on a lattice of size
    $L/a$, while the columns
    labeled $\Sigma(u,a/L)$ represent the value of the coupling on a
    lattice twice as large.
  Data in bold represent the new computations on larger lattices that
  require simulations on lattices up to $L/a=64$. 
This dataset allows a precise and accurate continuum extrapolation of
the step scaling function (see Figure~\ref{fig:fit_ssf}).}
  \label{tab:ssf_data}
\end{table}

\newpage

\begin{figure}
  \centering
  \includegraphics[width=\textwidth]{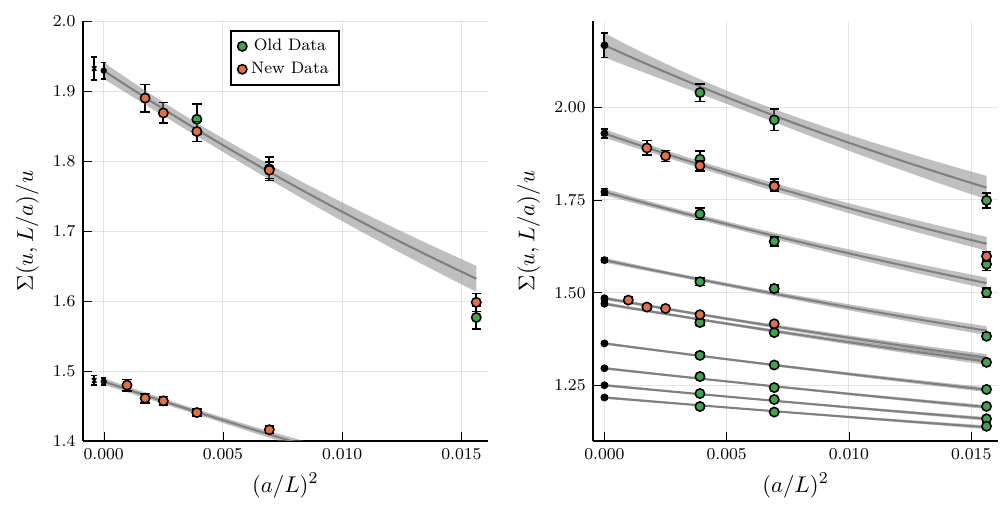}  
  \caption{Continuum limit of the step scaling function. 
    The dataset of table~\ref{tab:ssf_data} allows for a precise
    determination of the step scaling function in the continuum limit. 
    Left: In particular at two values of the coupling ($u=3.9490, 5.8673$)
    we have determined the lattice step scaling function $\Sigma(u,
    L/a)$ for lattice sizes up to $L/a =32$. 
   The continuum values are significantly improved in precision (the old continuum
   values are displayed with a cross), but central values have barely moved.
   Right: The full dataset, that includes both the new data and the
   old data~\cite{DallaBrida:2016kgh}. 
  }
  \label{fig:fit_ssf}
\end{figure}

\begin{figure}
  \centering
  \includegraphics[width=\textwidth]{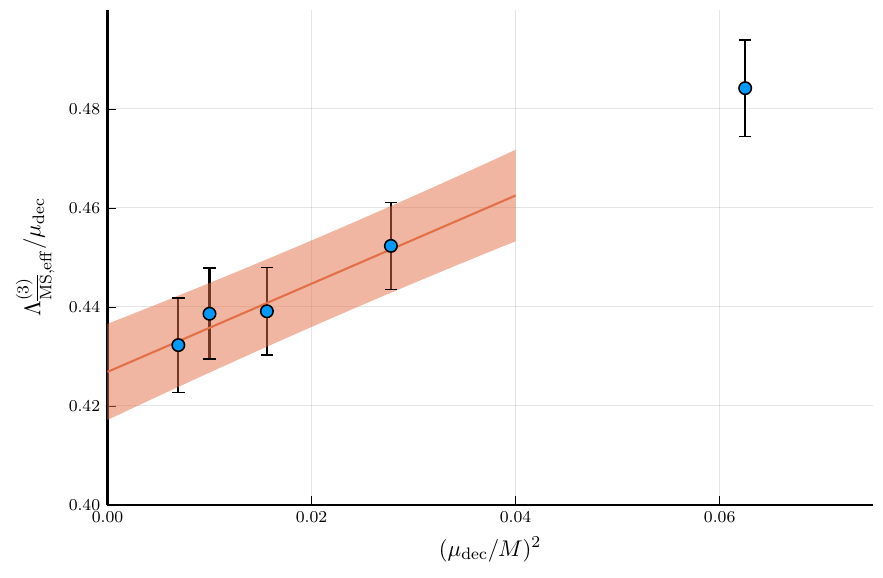}
  \caption{In the decoupling approach the intrinsic scale of QCD is
    extracted after matching $N_{\rm f} =3$ QCD and the pure gauge
    theory using three degenerate heavy quarks of mass $M$. 
    In the limit $M\to\infty$ the decoupling relation becomes
    exact. We performed simulations with quark masses in the range
    $M\approx 3 - 10$ GeV. This plot shows that the approach to infinite 
    mass (with corrections $\rmO(1/M^2)$)  agrees with the expectations 
    of an analysis based on effective field theory. The $M\to\infty$ limit 
    is obtained considering only masses $M\gtrsim 5\,\GeV$.}
  \label{fig:mtoinf}
\end{figure}

\begin{figure}
  \centering
  \includegraphics[width=\textwidth]{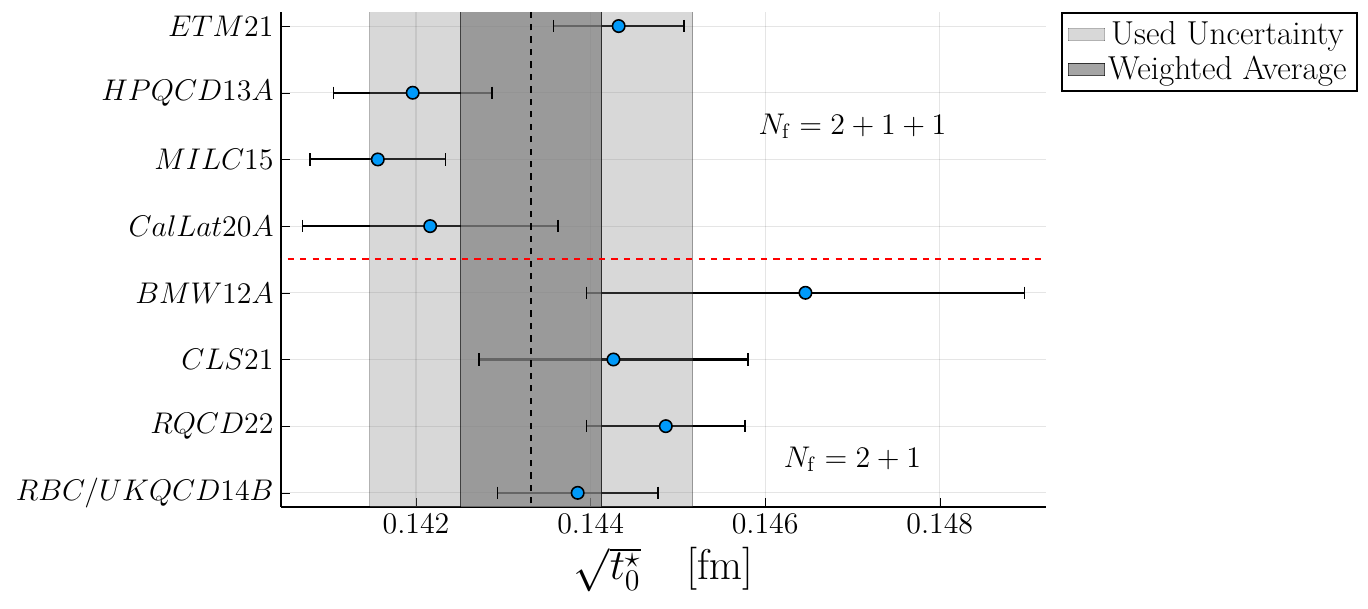}
  \caption{The length scale $\sqrt{t_0^\star}$ in fm using the ratio
    $\sqrt{t_0/t_0^\star}$ of eq.~(\ref{eq:t0t0star}) and the results for $\sqrt{t_0}$ from different collaborations~\cite{Miller:2020evg,Bazavov:2015yea,Dowdall:2013rya,Blum:2014tka,Alexandrou:2021bfr,Borsanyi:2012zs,Strassberger:2021tsu,RQCD:2022xux}.     
  Our conservative average includes the
  central values of  all precise results, and has an uncertainty
  almost two times larger than a weighted average. It thus takes the
  presently poor agreement into account. } 
\label{fig:murefs}
\end{figure}

\begin{table}
  \centering
  \begin{tabular}{lllcccccc}
  \toprule
  &&&\multicolumn{5}{c}{$z=M/\mu_{\rm dec}$} \\
  \cmidrule(lr){4-8}
  $c_{\rm M}$ &  $L/a$ & $6/\tilde g_0^2$ &$4$&$6$&$8$&$10$&$12$\\
  \cmidrule(lr){1-8}
-0.2     & 12 & 4.3019(16) & \textbf{5.235(20)} &         5.614(23)  &         6.107(19)  &                    & \\[0pt]
-0.2     & 16 & 4.4656(23) & \textbf{5.195(29)} & \textbf{5.498(31)} &         5.892(28)  &                    & \\[0pt]
-0.2     & 20 & 4.6018(24) & \textbf{5.259(31)} & \textbf{5.422(29)} & \textbf{5.757(33)} &                    &                6.481(33) \\[0pt]
-0.2     & 24 & 4.7166(25) & \textbf{5.258(33)} & \textbf{5.425(29)} & \textbf{5.678(38)} &         5.998(32)  &                6.281(30) \\[0pt]
-0.2     & 32 & 4.9000(42) & \textbf{5.255(47)} & \textbf{5.393(36)} & \textbf{5.558(33)} & \textbf{5.882(39)} & \textbf{       6.065(42)}\\[0pt]
-0.2     & 40 & 5.0497(41) &                    & \textbf{5.432(42)} & \textbf{5.562(40)} & \textbf{5.792(46)} & \textbf{       5.924(41)}\\[0pt]
-0.2     & 48 & 5.1741(54) &                    &                    & \textbf{5.502(48)} & \textbf{5.699(50)} & \textbf{       5.978(59)}\\[2pt]
  -0.2& $\infty$ &  &  5.258(28) & 5.347(22) & 5.479(31) & 5.669(40) & 5.780(51)\\[0pt]
  \cmidrule(lr){1-8}
  -0.1& $\infty$ &  &5.253(28) & 5.346(22) & 5.474(31) & 5.658(39) & 5.760(51)\\[0pt]
  -0.3& $\infty$ &  &5.262(28) & 5.349(22) & 5.484(31) & 5.682(39) & 5.799(51)\\[0pt]
  \bottomrule
\end{tabular}
	

  \caption{Massive couplings at finite $L/a$ corrected to the non-perturbative $\bg$ value
  with eq.~(\ref{eq:shift}) and parameter $c_\mathrm{M}=-0.2$. 
In the last three rows we list the continuum limits of the couplings
also considering $c_\mathrm{M}=-0.1, -0.3$. The reader can see that
the effect of the parameter $c_{\rm M}$ is not significant in the
values of the coupling in the continuum. 
  The continuum limits are taken according to eq.~(\ref{eq:global})
  with $\Gamma_{\rm eff} = \Gamma_{\rm eff}' = 
    0$ using only the data highlighted in bold face. These satisfy the
    cut $(aM)^2 \leq 0.16$.  
  Different choices of $\Gamma_{\rm eff}, \Gamma_{\rm eff}'$ are
  considered later (cf.~Figure \ref{fig:lam_variations}).}  
  \label{tab:bgshifts}
\end{table}

\begin{figure}[ht]
  \centering
  \begin{subfigure}{0.475\textwidth}
  \includegraphics[width=\textwidth]{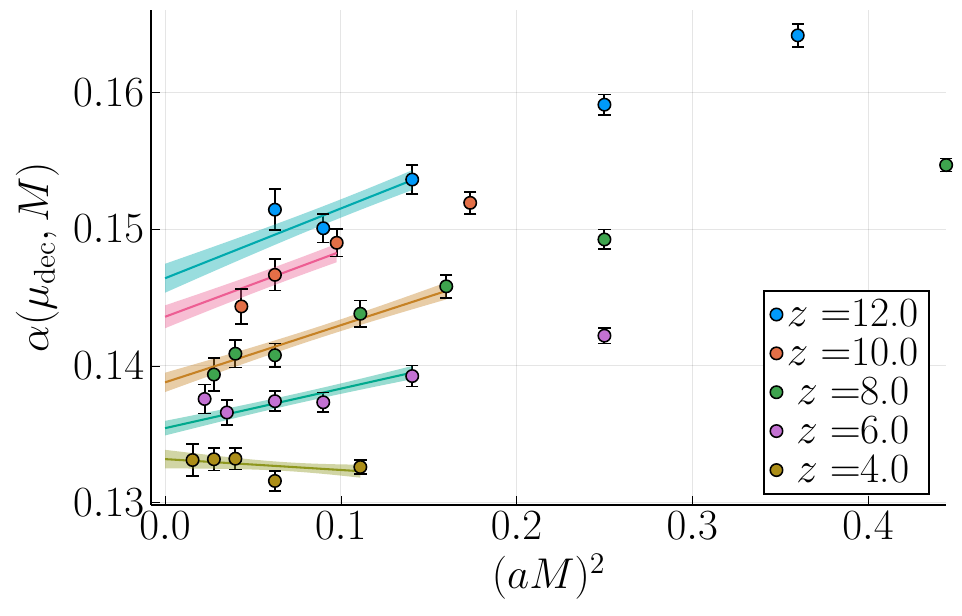}    
  \end{subfigure}
  \begin{subfigure}{0.475\textwidth}
  \includegraphics[width=\textwidth]{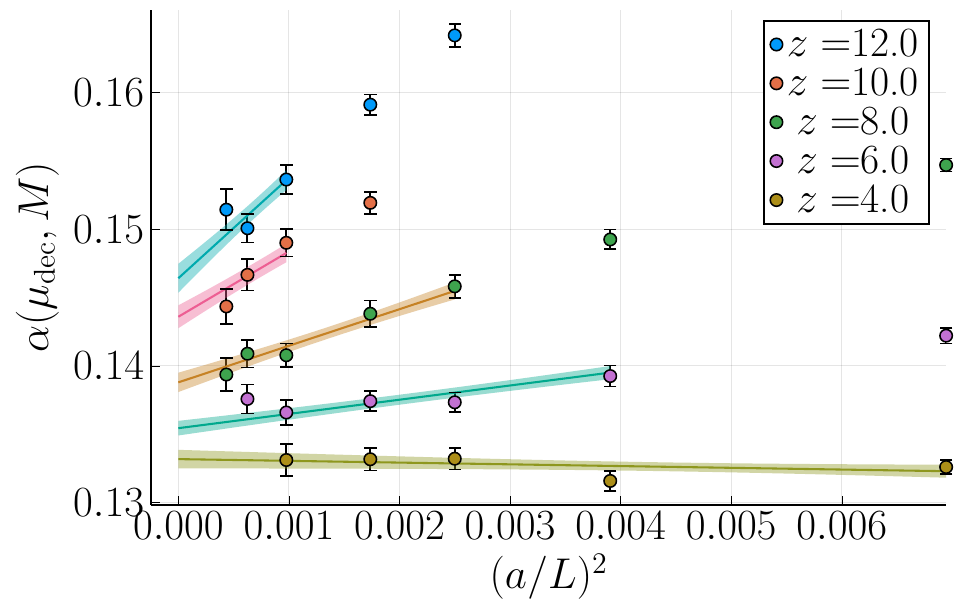}    
  \end{subfigure}

  \caption{Continuum extrapolation of our data for the massive coupling. 
  Our data, after being corrected with the non-perturbative value of
  $b_g$, agrees with expectations derived from an effective field
  theory analysis: most cutoff effects are 
  proportional to $(aM)^2$, with a constant slope.} 
  \label{fig:aextr030}
\end{figure}

\begin{figure}
  \centering
  \includegraphics[width=\textwidth]{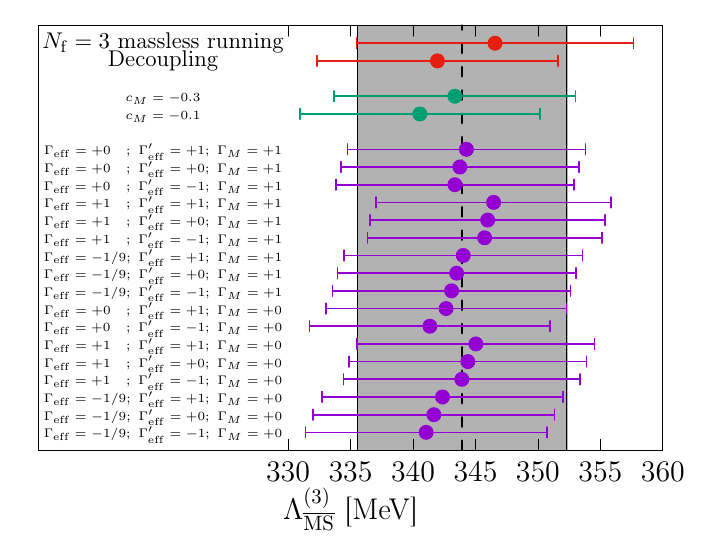}
  \caption{The two main results, using the massless running or the
    decoupling strategy, are displayed in red. Our final result (a
    weighted average of the previous two results) is displayed by
    vertical the grey band.
    Our result for $\Lambda^{(3)}_{\overline{\rm MS}}$ depends very
    little on the logarithmic corrections to the continuum
    extrapolation of the massive coupling Eq.~(\ref{eq:global}) as well as on 
    those in the large-mass extrapolation
    (\ref{eq:LargeMassExtrap});  these are all displayed by the purple points. 
  The dependence on the linear $\rmO(aM)$ shifts in $\Delta b_{\rm g}$ (green points) is even smaller
  (cf. 
  the parameter $c_{\rm M}$ in Eq.~(\ref{eq:shift})). 
}
  \label{fig:lam_variations}
\end{figure}

\begin{table}
  \centering
  \begin{tabular}{ccccc}
  \toprule
    $M/\mu_{\rm dec}$ & $\bar g_\mathrm{GFT}^2(\mu_{\rm dec},M)$ & $[\bar g_{\rm GF}^{(0)}(\mu_{\rm dec})]^2$ & $\rho^\mathrm{eff}$ & $\Lambda^\mathrm{eff}_{\overline{\rm MS}}$ [MeV] \\
    \midrule
    4  &  5.258(28)  &  4.048(18) &  0.4843(98) &  388.4(10.0) \\
    6  &  5.347(22)  &  4.103(15) &  0.4523(88) &  362.7(9.2) \\
    8  &  5.479(31)  &  4.184(19) &  0.4390(89) &  352.1(9.1) \\
    10 &  5.669(40)  &  4.299(24) &  0.4382(92) &  351.4(9.3) \\
    12 &  5.780(51)  &  4.364(31) &  0.4319(96) &  346.4(9.5) \\
    \midrule
    $M/\mu_{\rm dec}$ & $\hat\Gamma_{\rm M}$ & cut in $z$ & $\rho$ & $\Lambda_{\overline{\rm MS}}$ [MeV] \\
    \midrule
    $\infty$ & $0$ & $z\geq4$ & 0.4264(91) & 341.7(9.2)\\
    $\infty$ & $1$ & $z\geq4$ & 0.4323(89) & 346.3(9.1)\\
    $\infty$ & $0$ & $z\geq6$ & 0.4264(97) & 341.7(9.6) \\
    $\infty$ & $1$ & $z\geq6$ & 0.4286(95) & 343.5(9.5)\\
    \bottomrule
  \end{tabular}
  \caption{Values of the massive coupling in the continuum $\bar
    g^2_{\rm GFT}(\mu_{\rm dec}, M)$ and the corresponding values of the
    coupling in the pure gauge theory $[\bar g_{\rm GF}^{(0)}(\mu_{\rm  dec})]^2$ in the $c=0.3, T=L$ scheme. 
  We also quote the effective $\Lambda$-parameter, both in 
  units of $\mu_{\rm dec}$ and in
  physical
  units. 
Finally we also quote the extrapolations $M\to \infty$ of the latter. 
The couplings themselves diverge logarithmically in this limit.
}  \label{tab:lameff}
\end{table}

\begin{figure}
  \centering
  \includegraphics[width=\textwidth]{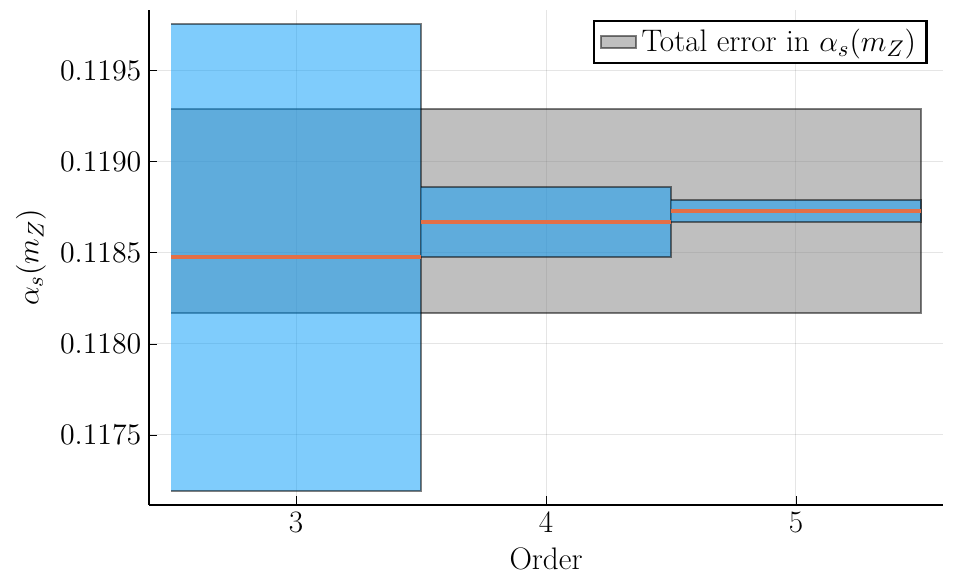}
  \caption{Translating the value of $\Lambda^{(3)}_{\overline{\rm MS}
    }$ into the value of the strong coupling requires crossing the
    charm and bottom quark thresholds using perturbation theory. 
    This plot shows the perturbative uncertainties at each order $n=3,
    4, 5$. 
    Orange horizontal lines represent $\alpha_s(mz)$ extracted
    from our central value $\Lambda^{(3)}_{\overline{\rm MS}}=343.9$
    MeV using different orders in perturbation theory. 
    The bands represent the perturbative uncertainties, estimated by
    the difference between the order $n$ and 
    the order $n-1$ result.
These perturbative estimates seem very conservative: 
for $n=3$ and $n=4$ the error bars are several times larger than
the actual difference between the order $n$ and the order $n+1$. 
Still the perturbative uncertainty for $n=5$ contributes less than 2\% in the
final error of $ \alpha_s(m_Z)$.}
  \label{fig:alphas_decv1}
\end{figure}

\FloatBarrier


\section{Supplementary material}

\subsection{Scale setting}
\label{sec:scale-setting}

Our result for the strong coupling depends on the value of a technical
scale $\mu_{\rm had}$ that has to be determined in physical
units. This scale is implicitly defined by specifying a precise value
for the massless coupling in 3 flavor QCD: $\bar g^2_{\rm GF}(\mu_{\rm had}) =
11.31$ (see~\cite{Bruno:2017gxd}). 
Ultimately, $\mu_{\rm had}$ has to be determined from a well measured
experimental quantity (like for instance the mass of the $\Omega$ baryon). 
In practice, it is convenient to introduce an intermediate technical length scale, 
$\sqrt{t_0^\star}$, derived from the gradient
flow~\cite{Luscher:2010iy}. 
This scale differs from the more common scale
$\sqrt{t_0}$ by the choice of quark masses: while $\sqrt{t_0}$ is
defined  for physical values of the quark masses, 
$\sqrt{t_0^\star}$ is defined for three degenerate quark masses $m_u
= m_d = m_s$ such that $12t_0 m_\pi^2 = 1.11$ (see
\cite{Bruno:2017gxd} for further details). Our estimate for
$\sqrt{t_0/t_{0}^\star}$, is based on \cite{RQCD:2022xux} but with an
enlarged error which also covers the central value from 
\cite{Strassberger:2021tsu}. Yet, the uncertainty on
\begin{equation}
  \sqrt{\frac{t_0}{t_0^\star}} = 1.0003(30)\,
  \label{eq:t0t0star}
\end{equation}
is negligible for our purposes.
This ratio allows us to determine $\sqrt{t_0^\star}$ from the results for $\sqrt{t_0}$ 
from different lattice collaborations. The estimates for the reference scale $\sqrt{t_0^\star}$
obtained in this way are displayed in Figure~\ref{fig:murefs}. 
We only show the computations that pass the averaging criteria set by
FLAG~\cite{FlavourLatticeAveragingGroupFLAG:2021npn}.\footnote{Note that CLS21~\cite{Strassberger:2021tsu} supersedes 
CLS16~\cite{Bruno:2017gxd} considered in FLAG~\cite{FlavourLatticeAveragingGroupFLAG:2021npn}.}
These computations were performed using a range of simulations that
allows for controlling the main sources of systematic uncertainties in lattice computations 
(i.e.~continuum limit, infinite volume extrapolation, chiral extrapolation, \dots). They also
use different experimental quantities as input (like the mass of the
$\Omega$ baryon, or leptonic decay rates  of Pions and Kaons).
Despite satisfying the FLAG criteria, the different data in
Figure~\ref{fig:murefs} show some tension. At present, there is no clear reason for this tension. Therefore 
we opt for a robust average. Our mean value for $\sqrt{t_0^\star}$ is given by the weighted average of all results, 
but we increase the error to include the central values of all computations which exert a pull of two or more 
on the weighted average.\footnote{The pull is defined as the contribution to the $\chi^2$ of the fit to a constant. 
Computations with a small pull either have a central value that agrees well with the mean or they have a large 
uncertainty and are therefore not so relevant in determining the mean.} 
As a result, we use 
\begin{eqnarray}
  \sqrt{t_0^\star} &=& 0.1434(19)\,{\rm fm}\,.
 \end{eqnarray}
Exploiting this result and the dimensionless combination 
  \begin{equation}
\label{eq:t0starmuhad}    
    \sqrt{t_0^\star}\,\mu_{\rm had} = 0.1456(11) \quad [0.7\%]\,,
  \end{equation}
that was accurately determined in~\cite{Bruno:2017gxd},
we can quote a value for the scale $\mu_{\rm had}$ in physical units
\begin{equation}
   \mu_{\rm had} = 200.5(3.0)\,{\rm MeV}\,.
\end{equation}
Despite the modest precision on this scale (1.5\%), its uncertainty is
subdominant in the final uncertainty of the strong coupling.



\subsection{The running coupling at low energies}
\label{sec:step-scaling-running}


Once the scale $\mu_{\rm had}$ is determined in physical units, one
needs to compute the running of the coupling from $\bar g^2_{\rm
  GF}(\mu_{\rm had}) = 11.31$, up to high energy scales, where
perturbation theory accurately describes QCD. 
The determination of the $\beta$ function follows the strategy
explained in section~\ref{sec:methods} and used in~\cite{DallaBrida:2016kgh}. 
First, we determine the lattice step scaling function in the GF scheme
\begin{equation}
  \Sigma(u, a/L) = \bar g^2_{\rm GF}(\mu/2)\Big|_{\bar g^2_{\rm GF}(\mu) = u}\,.
\end{equation}
Our data set includes estimates of $\Sigma(u, a/L)$ from lattices with
sizes $L/a=8, 12, 16$, at nine different values of $u$, as considered
in~\cite{DallaBrida:2016kgh}. Moreover we have new data~\cite{Fritzsch:2018yag,Sommer:2023gap,Conigli:2023rod,Conigli:2024lxq} 
much closer to the continuum, with $L/a=20, 24, 32$, at two specific
values of $u=3.949, 5.8673$. This data set allows us to determine
precisely the step scaling function in the continuum
\begin{equation}
  \sigma(u) = \lim_{a/L\to 0}\Sigma(u,a/L)\,.
\end{equation}
We use a linear functional form in $a^2$ for the continuum extrapolations. 
As explained in detail in~\cite{DallaBrida:2016kgh,Strassberger:2021tsu},
our data shows significant discretization effects. 
These are well described by a simple $a^2$ term, but different
subleading terms, e.g., $a^4$, can affect the result of the
extrapolations.
Therefore we opt to give less importance to the data at coarse lattice
spacing. We do so by considering a weight $[\Delta_i]^{-2}$ which includes
the statistical error $\Delta^{\rm stat}_i$ and a term reducing the weight of data further away from the continuum. 
It has the expected dominating $a^4$ contamination built in,
\begin{equation}
  \Delta_i^2=[\Delta^\mathrm{stat}_i]^2 + [\Delta^{\rm sys}_i]^2\,,\quad \Delta^{\rm sys}_i = 0.05\,\Sigma_i\, 
  \bigg(8\frac{a}{L}\bigg)^4\, \frac{u}{u_{\rm
        max}}\,. 
        \label{eq:genweights}
\end{equation}
The term $[\Delta^{\rm sys}_i]^2$ is negligible compared to our
statistical uncertainties at all values of $u$ for lattices with $L/a\ge 12$, but
it becomes dominant at large values of $u$ for $L/a=8$. As a result, in addition to
reducing the relevance of data far away from the
continuum, the uncertainties of our results
in the continuum are also increased. \Eq{eq:genweights} is 
unchanged compared to our previous study~\cite{DallaBrida:2016kgh}, 
but we now have data closer to the continuum. A comparison of our
previous extrapolations and those with the new data set 
show a considerable reduction of the error (cf. figure \ref{fig:fit_ssf}).
Central values are in total agreement. This validates our approach to perform the 
continuum extrapolations.

We parameterize the $\beta$-function as
\begin{equation}
  \beta(x) = - \frac{x^3}{P_N(x^2)}        \,,
\end{equation}
where $P_N(x^2)$ is a polynomial of degree $N$ in $x^2$, and 
extract its coefficients by fitting the continuum step scaling function data, $\sigma$, to
\begin{equation}
    \label{eq:logsiglat}
    \ln 2  = \int_{\sqrt{u}}^{\sqrt{\sigma(u)}} \frac{{\rm d} x}{\beta(x)}\,.
\end{equation}
Alternatively one can directly fit the lattice step scaling function $\Sigma(u,a/L)$ by
including the cutoff effects in the fitting ansatze as
\begin{equation}
    \label{eq:logsiglat}
    \ln 2 + \left( \frac{a}{L}  \right)^2 Q_{n_c}(u) = \int_{\sqrt{u}}^{\sqrt{\Sigma(u,a/L)}} \frac{{\rm d} x}{\beta(x)}\,, 
\end{equation}
where $Q_{n_c}(u)$ is a polynomial in $u$ of degree $n_c$ without constant term. 
We obtain good fits with $N>1$ and $n_c>1$.\footnote{See \cite{Bruno:2022mfy} for how one determines the 
	quality of fits with general weights such as the ones described above.}
We choose the values $N=n_c=2$ to quote our final result for the
$\beta$ function, whose continuum form reads
\begin{equation}
  \beta(x) = - \frac{x^3}{p_0 + p_1x^2 + p_2x^4}\,,
\end{equation}
with
\begin{equation}
  p_0 = 16.14035,\quad p_1 = 0.1817488,\quad p_2 = -0.01067877\,,	
\end{equation}
and covariance
\begin{equation}
  {\rm cov}(p_i,p_j) = \left(
  {\footnotesize
    \begin{array}{lll}
    +3.7345229609\times 10^{-1} &  -1.2847238140\times 10^{-1} &   +9.7506456742\times 10^{-3} \\ 
   -1.2847238140\times 10^{-1} &   +4.6710306855\times 10^{-2} &  -3.6595093583\times 10^{-3} \\ 
    +9.7506456742\times 10^{-3} &  -3.6595093583\times 10^{-3} &   +2.9433371344\times 10^{-4} \\ 
    \end{array}
  }
  \right)\, .
\end{equation}
This result allows us to compute some key ratios of scales. 
Specifically, defining
\begin{equation}
  \bar g^2_{\rm GF}(\mu_{\rm had}) = 11.31\,,\quad
  \bar g^2_{\rm GF}(\mu_{\rm dec}) = 3.949\,,\quad 
  \bar g^2_{\rm GF}(\mu_{0}/2) = 2.6723(64)\,,\quad 
\end{equation}
we obtain
\begin{equation}
  \frac{\mu_{\rm dec}}{\mu_{\rm had}} =        4.000(30)\,,\quad
  \frac{\mu_{0}}{\mu_{\rm dec}}  =        5.467(62)\,,\quad
  \frac{\mu_{0}}{\mu_{\rm had}}  =        21.86(33)\,.
\end{equation}
Given these ratios and the results of section~\ref{sec:scale-setting}, 
we can determine all these scales in physical units,
\begin{equation}
  \label{eq:lowscales}
  \mu_{\rm had} = 200.5(3.0)\, {\rm MeV}\,,\quad
  \mu_{\rm dec} = 802(13)\, {\rm MeV}\,,\quad
  \mu_{0}      = 4381(93)\, {\rm MeV}\,.
\end{equation}


\subsection{The strong coupling from $N_{\rm f} = 3$ QCD}
\label{sec:strong-coupling-from}

Given the knowledge of a hadronic scale $\mu_{\rm had}$ determined from the QCD
spectrum (see section~\ref{sec:scale-setting}), and a precise running of
the strong coupling from $\mu_{\rm had}\approx 200\,{\rm MeV}$ to $\mu_0\approx 4\, {\rm GeV}$ at
hand, it is time to match with perturbation theory and extract the
value of the strong coupling at high energy.

 One may be tempted to match with perturbation theory directly at $\mu_0$; 
 in fact perturbation theory is routinely used at these energy scales. 
This, however, would produce a result with very small statistical
uncertainties (below 0.2\% in $\alpha_s(m_Z)$), but with a difficult-to-estimate theoretical uncertainty. 
The defining feature of the strategy pursued by the ALPHA collaboration consists instead in pushing
the computation of the non-perturbative running of the coupling to much higher energies, up to the
electroweak scale, where the theoretical uncertainties associated with
the use of perturbation theory are negligible. 
The trade off is the larger statistical uncertainty that the determination of this
non-perturbative running carries. 

In~\cite{Brida:2016flw,DallaBrida:2018rfy} we determined the ratio
\begin{equation}
  \label{eq:lam3ovmu0}
  \frac{\Lambda^{(3)}_{\overline{\rm MS}} }{\mu_0} = 0.0791(19)\,.
\end{equation}
We recall that this result was obtained by matching with
perturbation theory at a scale $\mu_{\rm PT} \approx 70$ GeV. 
The convergence of the perturbative approach at high energy scales was
checked using three different renormalization schemes, and
perturbative uncertainties were estimated in many different ways,
including scale variation~\cite{DallaBrida:2018rfy}. 
Our result also includes a more accurate modeling of the $\mathcal
O(a)$ boundary effects~\cite{DallaBrida:2018rfy}.
Together with the result for $\mu_0$ (Eq.~(\ref{eq:lowscales})), we
get
\begin{equation}
  \label{eq:lam3nf3}
  \Lambda^{(3)}_{\overline{\rm MS} } = 346(11)\, {\rm MeV}\,.
\end{equation}
Compared with our previous result~\cite{Bruno:2017gxd}, the central value has moved by less
than half a standard deviation. 
This comes essentially from the change in the scale $\sqrt{t_0^\star}$. 
Despite the fact that the new
dataset reduces the uncertainty in 
${\mu_{0}}/{\mu_{\rm had}}$ by 20\%, and we also have a slight reduction of the error in
$\Lambda^{(3)}_{\overline{\rm MS} }/\mu_0$ (Eq.~(\ref{eq:lam3ovmu0})), our determination of
$\Lambda^{(3)}_{\overline{\rm 
    MS}}$ shows only a marginal improvement compared with our previous
computation. Nevertheless, this update is important for two reasons. First, our analysis techniques, 
and in particular our method for calculating the continuum limit of the step scaling function at low 
energy, have been validated. Second, it highlights the crucial point that our uncertainties are primarily 
governed by the statistical uncertainties on the running in the high energy regime. 
In conclusion, using a non perturbative approach for computing the running of the strong coupling 
up to the electroweak scale, even if it removes the theoretical uncertainty, 
results in a substantial statistical uncertainty. 
This is reducible through extensive computational efforts, but would
profit from an alternative strategy.



\subsection{Decoupling of heavy quarks}
\label{sec:deco-heavy-quarks}

As we commented, the uncertainty in $\Lambda^{(3)}_{\overline{\rm
    MS} }$ is dominated by the high energy running. 
The challenge is therefore to improve the statistical uncertainty in
computing the running coupling at high energies.
Here we use the alternative strategy based on the decoupling of heavy quarks
introduced in~\cite{DallaBrida:2019mqg} and applied in~\cite{DallaBrida:2022eua}. 
This approach achieves a 
high precision by shifting the step scaling computations from QCD to the pure gauge theory,
where much better precision can be achieved (see~\cite{DallaBrida:2019wur,Nada:2020jay}).

Let us summarize the decoupling strategy as described in~\cite{DallaBrida:2022eua}.
Using a massive scheme for the coupling, $\bar g^2_{\rm GF}(\mu, M)$,
where all $N_{\rm f} $ quarks are taken 
degenerate with renormalization group invariant (RGI) mass $M$,
this coupling tends to its pure gauge theory counterpart in the limit of large quark masses,
$M \gg \Lambda,\mu$.\footnote{We refer the reader to Section \ref{sec:mass-renorm} for a proper definition of $M$.}  
In formulas,
\begin{equation}
  \label{eq:dec_relation}
  \bar g^2_{\rm GF}(\mu, M) = [\bar g^{(0)}_{\rm GF}(\mu)]^2 +  \rmO(\Lambda^2/M^2,\mu^2/M^2)\,, 
\end{equation}
where the  power corrections of order $\Lambda^2/M^2$ and $\mu^2/M^2$ were explicitly written. The shorthand notation 
used in this equation hides the fact that for a given $\mu$ (which we shall set equal to $\mu_{\rm dec}$ in our numerical application), 
the pure gauge coupling $\gbar^{(0)}_{\rm GF}(\mu)$ depends on the mass $M$ through the matching of the $\Lambda$-parameters 
of the two theories, which reads,
\begin{equation}
	\label{eq:P}
	\Lambda_\msbar^{(0)}=P(\Lambda_\msbar/M)\,\Lambda_\msbar \,.
\end{equation}
As discussed in \cite{Athenodorou:2018wpk}, the function $P$ is perturbatively known to very high accuracy. 
By this we mean that the relation is known to a high order  in perturbation theory and 
the higher order corrections decrease very rapidly for the values of the masses of interest. 

In perturbation theory, the function $P$ can be computed as,
\begin{equation}
   \label{eq:Pdef}
   P(M/\Lambda_{\msbar}) = {\varphi^{(0)}_{\msbar}(g_\star \sqrt{C(g_\star)})\over\varphi^{}_{\msbar}(g_\star)}\,.
\end{equation}
In this equation $\varphi^{(0)}_{\msbar}(g)$ and $\varphi_{\msbar}(g)$ refer to the function (\ref{eq:varphi}) for $0$ and $3$ flavours, respectively, 
in	 the ${\msbar}$ scheme. The coupling  $g_\star=\bar g_\msbar(m_\star)$, where $m_\star$ is defined by the implicit equation
$m_\star=\overline{m}^{}_{\msbar}(m_\star)$ with $\overline{m}^{}_{\msbar}(\mu)$ the running quark mass in the 
$\msbar$ scheme at the scale $\mu$ (see section \ref{sec:mass-renorm}). Note that the coupling $g_\star$ is a function of $M/\Lambda_{\msbar}$ alone, i.e. 
$g_\star\equiv g_\star(M/\Lambda_{\msbar})$~\cite{Alexandrou:2021bfr,DallaBrida:2022eua}.
The function $C(g)$ relates the couplings in the 3 flavour and pure gauge theory in the $\msbar$ scheme as,
\begin{equation}
	   [\bar{g}^{(0)}_\msbar(m_\star)]^2 = C(\bar{g}^{}_{\msbar}(m_\star)) \bar{g}^2_{\msbar}(m_\star) \,,
\end{equation}
and it is known to 4-loop order in perturbation theory~\cite{Chetyrkin:2005ia,Schroder:2005hy,Gerlach:2018hen}.

Dividing eq.~(\ref{eq:P}) on both sides by $\mu_\mathrm{dec}$ yields \eq{eq:rhoeq},
which we rewrite here in a more precise form by including the power corrections
from \eq{eq:dec_relation},
\begin{equation}
  \rho
  P \left( z/\rho\right)   =
  \frac{\Lambda^{(0)}_{\overline{\rm MS}}}{\Lambda^{(0)}_\text{GF}} \times \varphi_{\rm GF}^{(0)}\left(\gbar_\mathrm{GF}(\mudec,M)\right) + \rmO(\Lambda^2/M^2,\mu^2/M^2)\,,
  \quad \rho\equiv \Lambda_\msbar/\mu_\mathrm{dec}\,. 
  \label{eq:central_decoup_w_powers}
\end{equation}
The scheme change $\Lambda^{(0)}_{\overline{\rm MS}} /
\Lambda^{(0)}_\text{GF} = 0.4981(17)$ \cite{DallaBrida:2017tru}
involves no approximation by  
virtue of \eq{e:LambdaRatio}; we  profit from the full accuracy of $P$
in the $\msbar$ scheme by working in terms of $\Lambda$-parameters.%
\footnote{We note that section \ref{sec:1loopMassiveGFcoupling} contains a perturbative analysis of the decoupling strategy 
based on eq.~(\ref{eq:central_decoup_w_powers}). This may provide the reader with a different and useful perspective on the 
general ideas presented in this and the following sections.}
 
In~\cite{DallaBrida:2022eua} we computed the massive coupling $\bar g^2_{\rm GF}(\mu_{\rm dec}, M)$ for values of the quark 
masses\footnote{A more precise definition of the massive coupling that we considered is given in the following sections.}
\begin{equation}
  M/\mu_{\rm  dec} \approx 4, 6, 8, 10, 12\,.
\end{equation}
The determination of $\bar g^2_{\rm GF}(\mu,M)$ requires a continuum
extrapolation of the corresponding lattice estimates at non-zero lattice spacing. 
These extrapolations are difficult. On the one hand, solid continuum extrapolations
require $aM \ll 1$. On the other hand, large values of $M$ are needed in order to have small 
$\rmO(1/M^2)$ corrections in eq.~(\ref{eq:central_decoup_w_powers}). 
This multi scale problem can only be solved by employing large
lattices in a finite volume scheme. These schemes avoid in fact the
introduction of any additional scale since $\mu=1/L$ 
(see~\cite{DallaBrida:2019mqg,DallaBrida:2022eua} for the details).

An additional problem in performing these continuum extrapolation is the presence
of linear $aM$ cutoff effects. The most critical ones to deal with originate from 
a term $\sim aM \tr F^2$ in the Symanzik effective theory, which describes the expansion of lattice observables 
in powers of the lattice spacing $a$ \cite{Symanzik:1983dc,Symanzik:1983gh,Luscher:1984xn,Sheikholeslami:1985ij,Luscher:1996sc},
accompanied by (non-integer) powers of $\alpha(1/a)$ \cite{Balog:2009yj,Balog:2009np,Husung:2019ytz,Husung:2021mfl,Husung:2022kvi}.
The $\sim aM \tr F^2$ term can be removed by a proper change of the bare coupling $g_0$ as one varies the quark mass~\cite{Luscher:1996sc}. 
More precisely, when computing the mass-dependence of an observable such as $\bar g^2_{\rm GF}(\mu,M)$, the quantity to keep fixed 
in order to have a fixed lattice spacing $a$ is $\tilde g_0^2 = g_0^2(1+b_{\rm g}(\tilde{g}_0^2) am_q)$ rather than $g_0$.
This guarantees that the $\rmO(aM)$ effects due to the $\sim aM \tr F^2$ term are absent in continuum extrapolations.\footnote{The Symanzik effective theory
also predicts a term $\sim aM^2\psibar\psi$. This, however, can be easily removed through a proper definition 
of the RGI quark mass $M$ (see section \ref{sec:mass-renorm} for more details).}

Until recently, the improvement parameter $b_{\rm  g}(\tilde{g}_0^2)$ was only known to leading order in perturbation theory~\cite{Sint:1995ch}. This 
approximation was used in our previous work~\cite{DallaBrida:2022eua}. Our error included an estimate of the effect that 
the difference between the (at the time unknown) non-perturbative value of $b_{\rm g}$ and its 1-loop approximation could have on the values of the coupling at finite 
lattice spacing. This estimate
dominated the uncertainty of the continuum extrapolation of the massive
couplings $\bar g^2_{\rm GF}(\mu, M)$. 
In particular, this source of uncertainty was systematic and could only be estimated through perturbative arguments. 
Below we shall rely on our recent non-perturbative determination of $b_{\rm
  g}$~\cite{DallaBrida:2023yka} in order to obtain precise continuum
extrapolations of the massive coupling $\bar g^2_{\rm GF}(\mu_{\rm dec}, M)$ free of this systematic uncertainty.

\subsubsection{Continuum limit of the massive couplings}
\label{sec:cont-limit-mass}

We start with a precise definition of the massive coupling. 
The Gradient Flow coupling is defined with
Schr\"odinger Functional boundary conditions as~\cite{Fritzsch:2013je}
\begin{equation}
  \bar g^2_{\rm GFT}(\mu, M) = \frac{1}{\hat{\mathcal N}(t, a/L)}
  t^2\langle E(t, x) \rangle\Big|_{T=2L,\, x_0=T/2,\,\sqrt{8t}=0.36\, L}\,,
  \label{eq:GFTcoupl}
\end{equation}
where $E(t,x)$ is the energy density of the flow fields and $\hat{\mathcal N}$ a normalization constant.
We focus on the preferred value  $c=0.36$ for $c=\sqrt{8t}/L$,
but we thoroughly checked that different choices, $c\in\{0.30, 0.33, 0.36, 0.39,
  0.42\}$, lead to completely equivalent results for $\Lambda_{\msbar}^{(3)}$.   
Note the  choice $T=2L$, where $T$ is the time extent of our space-time lattices. 
It defines the $\mathrm{x}=\rm GFT$ scheme, which is employed only for the massive coupling, while everywhere else,
e.g. in the definition of $\mu_\mathrm{dec}$, the $\mathrm{x}=\rm GF$ scheme 
with $c=0.3$ and $T=L$ is used. A non-perturbative matching of the two schemes will be discussed
below. The primary reason for having $T=2L$ for the massive coupling is that with this choice
boundary effects $\propto 1/M$ are suppressed to a level well below
our statistical uncertainties (see section~\ref{sec:boundary} for details).

\phantom{.}
\\[1ex]
{\bf Couplings on lines of constant physics}
\\[1ex]
The massive couplings are needed at fixed $L=1/\mudec$ and for several values of the 
RGI mass $M$. Furthermore, the simulations employ a certain resolution  $a/L$. The 
dimensionless quantities defining  $\mudec$ and $M$ are $\bar g_{\rm GF}^2(\mudec)=3.949$ and $z=M/\mudec=ML$, 
where it is understood that $\bar g_{\rm GF}^2(\mudec)$ is defined for $M=0$, i.e. for a bare subtracted mass $\mq=0$. 
As determined in \cite{DallaBrida:2022eua}, the bare couplings $\tilde g_0^2$ listed in \tab{tab:bgshifts} 
ensure $\bar g_{\rm GF}^2(\mudec)=3.949$. The necessary values of $\mq$ then follow from
$aM=z\times a/L$ and the ALPHA-collaboration's work on the non-perturbative quark mass renormalization,
which we summarize in \sect{sec:mass-renorm}. Given $a\mq$, one needs to
perform the massive simulations with 
bare couplings $g_0^2$ such that the improved bare coupling $\tilde g_0^2
= (1+b_\mathrm{g}(\tilde g_0^2) a\mq )\,g_0^2$ is kept fixed.
As discussed earlier, the improvement coefficient $\bg$ is needed to
cancel linear $a$-effects when connecting the theories at
different masses, but fixed lattice spacing.  For our range of $\tilde g_0^2$,
the non-perturbative $b_{\rm g}$ is given by \cite{DallaBrida:2023yka}
\begin{equation}
  \label{eq:bgnp}
  b_{\rm g}(\tilde g_0^2) = b_\mathrm{g}^{(1)} \tilde g_0^2 -0.0151 \tilde g_0^4 + 0.0424 \tilde g_0^6\,,\quad b_\mathrm{g}^{(1)}=0.036\,.
\end{equation}
However, our simulations were performed when only the 1-loop approximation 
$b_{\rm g} = 0.036 \tilde g_0^2 + \rmO(\tilde g_0^4)$ was available. 
We deal with this mismatch by correcting the data for $\bar g_{\rm GFT}^2(\mu_{\rm dec}, M)$ in Table~2 of~\cite{DallaBrida:2022eua} for
the change in the bare coupling due to the change in $b_{\rm g}$ via
\begin{equation}
  \label{eq:linaprox}
  \bar g_{\rm GFT, corrected}^2(\mu_{\rm dec}, M) = \bar g_{\rm GFT}^2(\mu_{\rm dec}, M)  - \frac{{\rm d} \bar g^2_{\rm GFT}(\mu_{\rm dec}, M)}{{\rm d} \tilde g_0^2} \Delta b_\mathrm{g}\,\tilde g_0^2\, am_q\,.
\end{equation}
The straightforward  $\Delta b_\mathrm{g}=b_\mathrm{g}(\tilde g_0^2) - b_\mathrm{g}^{(1)}\tilde g_0^2$ will be modified below. 

As discussed in detail in~\cite{DallaBrida:2022eua}, the derivative
\begin{equation}
  \frac{{\rm d} \gbar^2_{\rm GFT}(\mu_{\rm dec}, M)}{{\rm d} \tilde g_0^2} =
    \frac{\bar g_{\rm GFT}\beta^{(0)}_\mathrm{GFT}(\bar g_{\rm GFT})(1-\etaM(g_\star))}{\tilde g_0 \beta_0^{(3)}(\tilde g_0) }[1+R_\mathrm{z}+R_\mathrm{a}]\,,
    \label{eq:deriv1}
\end{equation}
is known with uncertainties that affect the corrected couplings only well below their statistical uncertainties. 
Here we shall only quote the final expression for this derivative, but encourage
the interested reader to consult the original references to understand its derivation (see appendix D
of~\cite{DallaBrida:2022eua} and references therein). We thus take
\begin{eqnarray}
  \beta^{(0)}_\mathrm{GFT}(x) & \approx & -k_0\, x^3\, ,\\
  1-\eta^{\rm M}(g_\star) &\approx & \frac{9}{11}\, ,\\
  \beta_0^{(3)}(x) &\approx & -0.054\, x^3\, ,\\
  R_{z} & \approx &k_0\frac{b}{z^2} \gbar^2_{\rm GFT}\, , \\                              
  R_{a} & \approx & \frac{p_1+p_2z^2}{\gbar_{\rm GFT}\beta_{\rm GFT}^{(0)}(\gbar_{\rm GFT})(1-\eta^{\rm M}(g_\star))} \left(\frac{a}{L}\right)^2 \,,
\end{eqnarray}
where for the case $c=0.36$ we have: $p_1 = -21.4, p_2=1.08, b = 10.54$, and $k_0 = 0.076$.

Beside the accuracy of the derivative (\ref{eq:deriv1}) itself, we also need to worry about the size of the quadratic (and higher order) terms in $\Delta b_\mathrm{g}$
which are neglected in the linear approximation (\ref{eq:linaprox}). We therefore want small $\Delta b_\mathrm{g}$ values. 
This can be achieved by adding to $\Delta b_{\rm g}$ a term which only modifies the $\rmO((aM)^2)$ terms in $\bar g_{\rm GFT, corrected}^2$, i.e. it is 
linear in $aM$,
\begin{equation}
  \label{eq:shift}
  \Delta b_\mathrm{g} = b_\mathrm{g}(\tilde g_0^2) - b_\mathrm{g}^{(1)}\tilde g_0^2 + c_\mathrm{M} \times (aM)\,.
\end{equation}
As long as the linear approximation (\ref{eq:linaprox}) remains accurate,
the coefficient $c_\mathrm{M}$ can be chosen at will. As anticipated, its effect is just an
additional cutoff effect of $\rmO((aM)^2)$ in the massive couplings, which does not affect the continuum values. 
For $c_\mathrm{M} = -0.2$ the shift in  $\bar g_{\rm GFT}$ is very small for  the data point
with $L/a=12$ and $z=6$. This choice results in very small shifts also for all other lattices 
that are relevant for the continuum extrapolation. 
The maximum value of $ |\Delta b_\mathrm{g}|$ is 0.03
and the maximum shift $|\bar g_{\rm GF, corrected}^2 - \bar g_{\rm
  GF}^2|$ is 0.1~. As a confirmation that the choice of $c_\mathrm{M}$
is not crucial, we repeated the entire analysis with
$c_\mathrm{M}=-0.1$ and $c_\mathrm{M}=-0.3$, see \tab{tab:bgshifts}.

\phantom{.}
\\[1ex]
{\bf Continuum extrapolation}
\\[1ex]
For the continuum extrapolation we consider the fit ansatz
  \begin{equation}
    \label{eq:global}
    \bar g^2_{\rm GFT}(\mu_{\rm dec},M_i) =
    \bar g^2(z_i) +
    p_1 [\alpha_s(a^{-1})]^{\hat \Gamma_{\rm eff}} {a^2\mu_{\rm dec}^2} +
    p_2 [\alpha_s(a^{-1})]^{\hat \Gamma'_{\rm eff}} (aM_i)^2 \,,
  \end{equation}
where $\bar g^2(z_i), p_1, p_2$ are fit parameters and $\alpha_s\equiv\alpha_{\overline{\rm MS}}^{(\Nf=3)}$. 
The fit parameters $\bar g^2(z_i)$ are the desired 
continuum couplings. The  fit function is a result of an analysis of the Symanzik effective theory in the limit of large $M$~\cite{DallaBrida:2022eua}. 
Dropping relative corrections of order $\rmO(1/M^2)$ in the effective theory reveals that there are 
   no cross-terms of the from 
  $a^2M\mu_{\rm dec}$ and that there are global coefficients 
  $p_1,p_2$ instead of individual terms $\tilde p_i a^2$ for each value of $z_i=M_i/\mu_\mathrm{dec}$.
Due to the expansion in $1/M$, the functional form \eq{eq:global} is only expected to describe data at
  large values of $M$. 
  For our case it is a very good approximation for $z = M/\mu_{\rm
    dec} \ge 4$. Indeed,  the slopes in
  the extrapolations in $(aM)^2$ in figure~\ref{fig:aextr030} (left)  vary little with $z$.

Table~\ref{tab:bgshifts} collects the data that is included in the global
fit, as well as the results of the 	continuum extrapolations. 
We also give the continuum values for several choices of the shift parameter
$c_\mathrm{M}$. As we can see from the results in the table, varying $c_{\rm M}$ in the range $-0.1, -0.3$ around our preferred value $-0.2$
results in variations of the continuum values of at most $1/3$ of the statistical uncertainties. 

The effect of the non-perturbative knowledge of $b_\mathrm{g}$ is crucial for 
the accuracy of our  continuum limits. 
Compared with our previous results of~\cite{DallaBrida:2022eua}, our continuum data is now between three
to four times more precise.
It is also worth noting that after the correction the data is much more
linear in $a^2$. Judging by the quality of the fit alone, we could in fact extend the fitting range to include data
up to $(aM)^2 \approx 0.3$ (and thus reduce 
the statistical uncertainties in the continuum even further).  We however decided to quote
as final result the fit that only includes data with $(aM)^2 < 0.16$. This
approach is conservative, since it leads to larger
uncertainties and entirely compatible central values. 
In addition, this choice keeps the effect of the logarithmic corrections in
the continuum extrapolation under control. 
Varying the corresponding parameters in the relevant range: $\Gamma_{\rm eff} \in [-1,1]$ and $\Gamma_{\rm
 eff}'\in [-1/9,1]$, the effect that we observe is typically of about the same size as the statistical errors on the
continuum values. As we shall discuss later, this effect is subdominant in
$\Lambda^{(3)}_{\overline{\rm MS} }$ and at most of the order of 10\% of the
statistical uncertainties on $\Lambda^{(3)}_{\overline{\rm MS}}$ (cf. Figure \ref{fig:lam_variations}).

In conclusion, we are able to obtain precise values for the massive
coupling in the continuum, where the choice of different ans\"atze for the continuum
extrapolations results in values for $\Lambda^ {(3)}_{\overline{\rm
    MS}}$ that vary much less than the statistical fluctuations. 

\subsubsection{The determination of $\Lambda^{(3)}_{\overline{\rm MS} }$}
\label{sec:LambdaFromDec}

In our strategy, \eq{eq:central_decoup_w_powers}, $\Lambda^{(3)}_\msbar$ is determined via the pure gauge running function 
$\varphi_{\rm GF}^{(0)}(g)$ in the GF scheme. On the other hand, the systematic errors due to boundary effects 
in the Schr\"odinger Functional are much easier to control for the GFT coupling, \eq{eq:GFTcoupl}. The idea is therefore 
to first apply the decoupling relation, \eq{eq:dec_relation}, to the GFT coupling, and in a second step switch to the GF 
scheme through the (pure gauge theory) function
\begin{equation}
	\gbar^{(0)}_{\rm GF}(\mu)=\chi_{0.36}(\gbar^{(0)}_{\rm GFT}(\mu))\,.
\end{equation}
Note that the scheme switch also involves the change in the value of $c$ from 
$c=0.36$ (GFT) to $c=0.30$ (GF). (Other values for $c$ in the GFT scheme have been considered 
for consistency checks.)

The precise form in which we apply the relation \eq{eq:central_decoup_w_powers} is thus the following,
\begin{equation}
  \rho^\mathrm{eff}
  P \left( z/\rho^\mathrm{eff}\right)   =
  \frac{\Lambda^{(0)}_{\overline{\rm MS}}}{\Lambda^{(0)}_\text{GF}} \times \varphi_{\rm GF}^{(0)}\left(\chi_{0.36}(\gbar_\mathrm{GFT}(\mudec,M)\right) \,,
  \quad \rho^\mathrm{eff}\equiv \Lambda_\msbar^\mathrm{eff}/\mu_\mathrm{dec}\,,
  \label{eq:central_decoup_eff}
\end{equation}
where we absorbed all power corrections of $\rmO(1/M^2)$ into an effective $\Lambda$-parameter, $\Lambda_{\overline{\rm MS}}^{\rm eff}$. 
The latter needs to be extrapolated for $M\to\infty$ in order to extract the desired result, i.e.
\begin{equation}
	\Lambda_\msbar = \Lambda_\msbar^\mathrm{eff}+\rmO(1/M^2)\,.
\end{equation}
We shall now turn to the details of the numerical analysis taking the
continuum values for the massive coupling as input (cf. table~\ref{tab:lameff}).

The functions $\varphi_{\rm GF}^{(0)}$ and $\chi_{0.36}$ can be determined from the results of~\cite{DallaBrida:2019wur} (see also appendix B of~\cite{DallaBrida:2022eua}).
Here we list accurate numerical representations valid for the relevant couplings. The function $\chi_{0.36}$ is implicitly defined by its inverse function
  $\chi^{-1}_{0.36}$ via the expression
\begin{equation}
	[\chi_{0.36}^{-1}(g)]^2=\frac{g^2}{1+g^2 P^{0.36}_{n_p}(g^2)}\,,
 \end{equation}
 where $P^{0.36}_2$ is a second degree polynomial with
      coefficients
 \begin{eqnarray}
 p_0 &=&  -7.8231\times 10^{-2}\, , \\ 
 p_1 &=&  \phantom{-}8.0579\times 10^{-3}\, , \\ 
 p_2 &=&  -6.8506\times 10^{-4}\, , 
 \end{eqnarray}
 and covariance
      \begin{equation}
        {\rm cov}(p_i,p_j) = \left(
  {\footnotesize
          \begin{array}{ccc}
 +3.07343869\times 10^{-4} &-1.35537879\times 10^{-4} &+1.48252105\times 10^{-5} \\
 -1.35537879\times 10^{-4} &+6.01436188\times 10^{-5} &-6.58280673\times 10^{-6} \\
 +1.48252105\times 10^{-5} &-6.58280673\times 10^{-6} &+7.24967403\times 10^{-7} \\
          \end{array}
          }
        \right)\, .
      \end{equation}
      The polynomial is accurate for couplings $g^2
      \in [3.8, 5.1]$.

The function $\varphi_{\rm GF}^{(0)}(g)$ defined in \eq{eq:varphi} determines the $\Lambda$-parameter in
the pure gauge theory as a function of $g$ in the GF scheme. We use the numerical representation
\newcommand{\gsw}{g_\mathrm{sw}}
\begin{eqnarray}
  \varphi_{\rm GF}^{(0)}(g) &=& f_0 \\
                   &\times&
                            \nonumber
 \exp \left\{ \frac{p_0}{2} \left( \frac{1}{\gsw^2} - \frac{1}{g^2} \right)
    + \frac{p_1}{2}\log \frac{\gsw^2}{g^2}
    + \frac{p_2}{2} \left( \gsw^2 - g^2 \right)
    + \frac{p_3}{4} \left( \gsw^4 - g^4 \right)
  \right\}\,,
\end{eqnarray}
with parameters
 \begin{align}
 \gsw^2 = 4\pi \times 0.2\,, \quad 
   f_0 = 0.2658(36)\,, \quad
   p_0 = 14.93613381 \,,
   \\
 	\quad 
 	p_1 =  -1.03947429 \,,
 	\quad
 	p_2 = 0.18007512 \,,
 	\quad
 	p_3 = -0.01437036 \,,
 \end{align}
and covariance
\begin{multline}
 	{\rm cov}(p_i,p_j) = \\[1ex]
 	\left(
          {\footnotesize
          \begin{array}{cccc}
 		+5.24669327\times 10^{\text{-1}} &   -3.26120586\times 10^{\text{-1}} &  +6.03484522\times 10^{\text{-2}} &   -3.33454413\times 10^{\text{-3}}\\
 		-3.26120586\times 10^{\text{-1}} &   +2.07627940\times 10^{\text{-1}} &  -3.91082685\times 10^{\text{-2}} &   +2.19046893\times 10^{\text{-3}}\\
 		+6.03484522\times 10^{\text{-2}} &  -3.91082685\times 10^{\text{-2}} &   +7.47684098\times 10^{\text{-3}} &  -4.23948184\times 10^{\text{-4}}\\
 		-3.33454413\times 10^{\text{-3}} &   +2.19046893\times 10^{\text{-3}} &  -4.23948184\times 10^{\text{-4}} &   +2.42972883\times 10^{\text{-5}}\\
          \end{array}
          }
	\right)\,.
\end{multline}
These functions are applied to the second column of \tab{tab:lameff} to determine columns three and four.

The $z\to\infty$ extrapolation of $	\rho^\mathrm{eff}$  is carried out by a fit to the leading behavior
in the large mass expansion,
\begin{equation}
	\rho^\mathrm{eff} = \rho + s\, [\alpha_s(m_\star)]^{\hat \Gamma_{\rm M}} \frac{1}{z^2}\,,
	\label{eq:LargeMassExtrap}
\end{equation}
where, we recall, $m_\star$ is defined by $\mbar_\msbar(m_\star) = m_\star$, and we vary $\hat \Gamma_{\rm M} = 0,1$. The results are found in 
\tab{tab:lameff}.
Taking the more conservative fits to the data with $z\geq6$ ($\hat{\Gamma}_{\rm M}=0$) we obtain
\begin{equation}
	\rho =0.4264(97)\,.
\end{equation}
The corresponding large mass extrapolation is shown in Figure \ref{fig:mtoinf}. 

We end this section with a remark on the different systematics that affect our computation. 
Despite the fact that we work in a finite volume, using very small
values of the lattice spacing ($a^{-1} \approx 10-50$ GeV, with
$\alpha_s(a^{-1})$ varying very little), the continuum values of the
massive coupling are affected by the choice of $\hat{\Gamma}_{\rm eff},\hat{\Gamma}'_{\rm eff}$ (cf.~eq.~(\ref{eq:global})). 
This illustrates how difficult continuum extrapolations can be. 
Fortunately, our final goal is not the determination of the massive
coupling in the continuum itself, but rather the determination of
$\Lambda^{(3)}_{\overline{\rm MS}}$. 
As figure~\ref{fig:lam_variations} shows, the effect on $\Lambda^{(3)}_{\overline{\rm MS}}$ 
due to the change in the coupling values for different continuum extrapolations is very mild. 
Logarithmic corrections in the $M\to\infty$ extrapolation 
(cf. the parameter $\hat{\Gamma}_{\rm M}$ in eq.~(\ref{eq:LargeMassExtrap})) 
also have little effect on $\Lambda^{(3)}_{\overline{\rm MS} }$. 
All in all, our decoupling strategy has allowed us to extract $\Lambda$ with an
error dominated by statistical uncertainties.


\subsection{The strong coupling}
\label{sec:strong-coupling}

The two strategies that we employed for extracting $\Lambda^{(3)}_{\overline{\rm MS}}$
are largely independent, and are affected by very different systematics. 
First we quote the results of $\Lambda^{(3)}_{\overline{\rm MS} }$ in
units of $\sqrt{t_0}$ determined using each strategy
\begin{eqnarray}
   \text{Massless running: } \sqrt{8t_0}\times \Lambda^{(3)}_{\overline{\rm MS}} = 0.712(21) \,, \\
   \text{Decoupling: } \sqrt{8t_0}\times \Lambda^{(3)}_{\overline{\rm MS}} = 0.703(18)\,.
\end{eqnarray}
Using our estimate for $\sqrt{t_0}$ (cf. section~\ref{sec:scale-setting}) we obtain:
\begin{eqnarray}
   \text{Massless running: } &\Lambda^{(3)}_{\overline{\rm MS}} = 347(11)\, {\rm MeV}\,, \\
   \text{Decoupling: }  &\Lambda^{(3)}_{\overline{\rm MS}} = 341.9(9.6)\,  {\rm MeV}\,,
\end{eqnarray}
The results show a good agreement, with a correlation between the two
estimates of $\Lambda^{(3)}_{\overline{\rm MS} }$ of $6.5\times
10^{-5}$. Since the uncertainties are dominated by statistics and have
very different origin, an average is justified. A fit to a constant
leads to our final result 
\begin{eqnarray}
        \label{eq:lam3avg}
  \sqrt{8t_0}\times \Lambda^{(3)}_{\overline{\rm MS}} &=& 0.707(15)  \\
  \Lambda^{(3)}_{\overline{\rm MS} } &=& 343.8(8.4)\, {\rm MeV}\,.
\end{eqnarray}

All that is left to do now is to convert the result for
$\Lambda^{(3)}_{\overline{\rm MS} } $ into a value for the strong coupling. 
This is performed in two steps:\\

\begin{enumerate}
\item With $m_{\rm c}^\star =       1275.0(5.0)$ MeV and $m_{\rm b}^\star =
4171(20)$ MeV as values for the quark
masses~\cite{FlavourLatticeAveragingGroupFLAG:2021npn}, we use
perturbation theory to cross the charm and bottom thresholds:	
\begin{equation}
  \Lambda^{(5)}_{\overline{\rm MS} } = P_{4,5}(M_{\rm b}/\Lambda^{(5)}_{\overline{\rm MS}})
  \times P_{3,4}(M_{\rm c}/\Lambda^{(4)}_{\overline{\rm MS}}) \times \Lambda^{(3)}_{\overline{\rm MS} } \,.
\end{equation}
In this expression the $P_{N_{\rm f}, N_{\rm f}+1}$ factors are the ratios of $\Lambda$
parameters computed in perturbation theory after the decoupling of a single
quark of mass $M$,
\begin{equation}
  P_{N_{\rm f}, N_{\rm f}+1}(M/\Lambda^{(N_{\rm f} +1)}_{\overline{\rm MS}}) = \frac{\Lambda^{(N_{\rm f} +1)}_{\overline{\rm MS}} }{\Lambda^{(N_{\rm f})}_{\overline{\rm MS}} }\,.
\end{equation}
\item Using the experimental input $m_Z = 91188.0(2.0)$ MeV~\cite{ParticleDataGroup:2024cfk} one
determines $\alpha_s(m_Z)$ by solving the equation
\begin{equation}
  \frac{\Lambda^{(5)}_{\overline{\rm MS} } }{m_Z} = \varphi_{\overline{\rm MS}}^{(N_{\rm f}=5)}\left(  \sqrt{\alpha_s(m_Z)\times (4\pi)}\right)\,.
\end{equation}
\end{enumerate}
Both steps rely on perturbation theory and it is important to
estimate the effect of the truncation of the perturbative series on the final result. 

The second step carries negligible perturbative uncertainties thanks to the fact that $m_Z$ is a very high
energy scale and the $\beta$-function that enters the function
$\varphi_{\overline{\rm MS} }^{(N_{\rm f} =5)}$ is known to
5-loop order. The evaluation of the factors $P_{3,4}, P_{4,5}$, on the other hand,  requires to
use perturbation theory at relatively low energy scales (the charm and
bottom quark masses). 
These factors are also known in perturbation theory to 5-loop accuracy. 
The 3-, 4-, and 5-loop order terms of the product $P_{3\to5}=P_{3,4}\times P_{4,5}$
give contributions to $\alpha_s(m_Z)$ of $128\times 10^{-5}, 19\times 10^{-5}$, and $6\times 10^{-5}$, respectively.
This shows that the perturbative series is well behaved, with
fast decreasing contributions as the perturbative order is increased
(see figure~\ref{fig:alphas_decv1}).  
We use the last term in the series as an estimate of the perturbative truncation errors. 
As a check we tried several variants of scale variations by a factor of two, using the explicit matching formulae from
\cite{Chetyrkin:2005ia,Schroder:2005hy}, for either the running mass in the $\msbar$ scheme at a
given scale ($3\, \GeV$ for charm, $4-5\, \GeV$ for the bottom-quark), or by varying the ratios $\mu/m_\star$ 
(up and down for bottom and only up for the charm quark).
In the first case, a scale-variation up and down, for both the coupling
and the running quark mass, leads to uncertainties in $P_{3,4}$ and $P_{4,5}$ 
significantly smaller than our estimates.
The only effect somewhat exceeding our estimate was found in $P_{4,5}$ when varying $\mu/m_\star$ for the bottom quark,
and corresponds to $\text{($1.17$ vs.~$0.84$)} \times 10^{-5}$ in $\alpha_s(m_Z)$.

The factor $P_{3\to 5}$ also has non-perturbative contributions. 
These are more difficult to estimate. 
In~\cite{Athenodorou:2018wpk} the simultaneous decoupling of 2 quarks
was studied in detail. 
We use their estimate (0.2\% in $\Lambda$ per quark at the charm
mass) with an increased uncertainty (due to a potential difference between
the model $P_{0,2}$ and the physical case of QCD that is dominated by
$P_{3,4}$), to add
an additional 0.5\% uncertainty to $\Lambda^{(3)}_{\overline{\rm MS} }$.
These steps lead to our final value
\begin{equation}
  \label{eq:alphas}
    \alpha_s(m_Z) = 0.11873(56) \qquad [0.47\%]\,.
\end{equation}
Our result is very precise, still the uncertainties are dominated by
the uncertainty in $\Lambda^{(3)}_{\overline{\rm MS} } $ (eq.~(\ref{eq:lam3avg})). 
The two main sources of uncertainty are the running in the pure gauge
theory and the scale $\sqrt {t_0}$, each contributing approximately
25\% to the error squared in $\alpha_s(m_Z)$. The massless running at high
energies (eq.~(\ref{eq:lam3ovmu0})) contributes 16\% of the error
squared, and the dimensionless ratio $\sqrt{t_0^\star}\mu_{\rm had}$
(eq.~(\ref{eq:t0starmuhad})) contributes 10\%. 
The rest of the uncertainty in our final result comes from the
statistical uncertainties in the determination of the massive couplings. 
Note that most of the sources of uncertainty are statistical. In
particular both the non-perturbative and perturbative truncation
errors in $P_{3\to5}$ contribute each less than 2\% to the total error squared.
The perturbative uncertainties from the truncation of the function 
$P$ entering the decoupling relation, eq.~(\ref{eq:central_decoup_eff}), are even smaller,
and amount to only 1.7\% of the final error squared of $\alpha_s(m_Z)$. 


\subsection{Mass renormalization}
\label{sec:mass-renorm}
  
In the decoupling strategy, the continuum limit is taken at some prescribed values of the renormalization-group invariant 
(RGI) quark mass, $z=M/\mu_{\rm  dec}=4,6,\ldots,12$, measured in units of the scale $\mu_{\rm  dec}$ 
at which we apply decoupling (cf.~Section \ref{sec:deco-heavy-quarks}). On the lattice, this is achieved by a proper tuning of the bare 
quark-mass parameter $m_0$ appearing in the lattice Lagrangian as a function of the lattice cutoff $1/a$. 
More precisely, for the lattice discretization employed in this work, the relation between the RGI and bare quark mass 
reads%
\footnote{We recall that $\tilde{g}^2_0$ stands for the so called improved bare coupling (cf.~Section \ref{sec:deco-heavy-quarks})
and a given value of $\tilde{g}^2_0$ corresponds to a particular value of the lattice cutoff $1/a$.}
\begin{equation}
	M=Z_{\rm M}(\tilde{g}_0^2)\widetilde{m}_{\rm q}\,,
	\qquad
	\widetilde{m}_{\rm q}=m_{\rm q}(1+b_{\rm m}(\tilde{g}_0^2) am_{\rm q})\,,
	\qquad
	m_{\rm q}=m_0-m_{\rm crit}\,.
\end{equation}
In this equation, $Z_{\rm M}(\tilde{g}_0^2)$ is the renormalization factor that relates the so called improved bare subtracted quark
mass $\widetilde{m}_{\rm q}$ and the RGI mass $M$. It only depends on the value of the lattice cutoff through $\tilde{g}_0^2$, 
while, as $M$ itself, it is renormalization scheme and scale independent. The coefficient function $b_{\rm m}(\tilde{g}_0^2)$ 
appearing in $\widetilde{m}_{\rm q}$ is analogous to the function $b_{\rm g}(\tilde{g}_0^2)$ entering the definition of the 
improved bare coupling $\tilde{g}^2_0$: if properly tuned, it allows for the removal of $\rmO(a)$ discretization 
errors in the definition of the renormalized quark mass~\cite{Luscher:1996sc}.%
\footnote{We note that differently from ref.~\cite{Luscher:1996sc}, we take as an argument of the function $b_{\rm m}$ the improved bare coupling 
$\tilde{g}_0^2$ instead of $g_0^2$. The difference between these two choices amounts to an $\rmO((am_{\rm q})^2)$ effect in $M$.}
 Given $b_{\rm m}(\tilde{g}_0^2)$, the improved mass $\widetilde{m}_{\rm q}$ 
is related to the subtracted quark mass $m_{\rm q}$ by a quadratic equation. The bare mass $m_{\rm q}$ is defined in terms 
of the parameter $m_0$ that controls the simulations and the critical mass parameter $m_{\rm crit}\equiv m_{\rm crit}(g_0^2)$, 
which identifies the value of $m_0$ for which (up to discretization errors) the quarks are massless.

\begin{table}[h]
	\centering
	\begin{tabular}{lllll}
		\toprule
		$L/a$& $\beta$  & $am_{\rm crit}$        & $Z_{\rm m}$      & $b_{\rm m}$     \\
		\midrule
		$12$ & $4.3020$ & $-0.323417(38)$  & $1.6882(72)$ & $-0.42(20)$  \\
		$16$ & $4.4662$ & $-0.312928(23)$  & $1.7252(80)$ & $-0.50(12)$  \\
		$20$ & $4.5997$ & $-0.304289(24)$  & $1.739(10)$  & $-0.47(14)$  \\
		$24$ & $4.7141$ & $-0.296941(14)$  & $1.770(11)$  & $-0.51(10)$  \\
		$32$ & $4.9000$ & $-0.285427(12)$  & $1.813(17)$  & $-0.619(48)$ \\
		$40$ & $5.0671$ & $-0.275473(11)$  & $1.807(19)$  & $-0.50(10)$  \\
		$48$ & $5.1739$ & $-0.2693605(82)$ & $1.823(22)$  & $-0.528(73)$ \\
		\bottomrule
	\end{tabular}
	\caption{Results for $am_{\rm crit}$, $Z_{\rm m}\equiv Z_{\rm m}^{\rm SF}(g_0^2,a\mu_{\rm  dec})$, and $b_{\rm m}$, for different values
		of $L/a$ and $\beta=6/g_0^2$. See the main text for any unexplained notation.}
	\label{tab:MassRenorm}
\end{table}

In ref.~\cite{DallaBrida:2022eua}, the functions $Z_{\rm M}(\tilde{g}_0^2)$, $b_{\rm m}(\tilde{g}_0^2)$, and $m_{\rm crit}(g_0^2)$, have been determined for 
the values of the lattice spacing of interest from  
the results for $Z_{\rm m}$, $b_{\rm m}$, and $am_{\rm crit}$  collected
in Table \ref{tab:MassRenorm}.  The renormalization factor $Z_{\rm M}$ is determined via the introduction of an intermediate, mass-independent, 
renormalization scheme $r$ for the quark mass, according to the equation
\begin{equation}
	\label{eq:ZM}
	Z_{\rm M}(\tilde{g}_0^2)= \bigg({M\over \overline{m}_{r}(\mu_{\rm  dec})}\bigg) Z^{r}_{\rm m}(\tilde{g}_0^2,a\mu_{\rm  dec})\,.
\end{equation}
In this equation, we denoted with $Z^{r}_{\rm m}(\tilde{g}_0^2,a\mu_{\rm  dec})$ the renormalization constant that defines the 
corresponding renormalized, running, quark mass,
\begin{equation}
	\overline{m}_{r}(\mu_{\rm dec})=\lim_{a\to 0} Z^{r}_{\rm m}(\tilde{g}_0^2,a\mu_{\rm  dec})\widetilde{m}_{\rm q}\,,	
\end{equation}
in the renormalization scheme $r$ and at the renormalization scale $\mu=\mu_{\rm  dec}$. 

In order to compute the first factor on the r.h.s. of eq.~(\ref{eq:ZM}), one needs to determine, in the continuum limit, the 
RG evolution of the quark mass $\overline{m}_{r}(\mu)$ from the scale $\mu_{\rm dec}$, up to infinite energy. The RG running 
of the quark mass is encoded in the function $\varphi_{r,\rm m}(\bar{g}_s)$ that defines the RGI quark mass through,%
\footnote{For the ease of notation we label the function $\varphi_{r,\rm m}(\bar{g}_s)$ only with the scheme $r$ defining
the renormalized quark masses, although implicitly its form depends also on the scheme $s$ chosen for its argument, the gauge 
coupling. The same is true for the $\tau$-function. On the other hand, note that if either of these functions is expressed as a function 
of the renormalization scale $\mu$ in physical units, rather than the coupling $\bar{g}^2_s$, they only depend on $r$ (cf.~e.g.~ref.~\cite{Campos:2018ahf}).} 
\begin{gather}
	  \label{eq:M}
	  M=\overline{m}_r(\mu)\varphi_{r,\rm m}(\bar{g}_s(\mu))\,,\\[2ex]
	  \label{eq:varphim}
	  \varphi_{r,\rm m}(\bar{g}_s)=	  
	  \left[2b_0\bar{g}_s^2 \right]^{-\frac{d_0}{2b_0}}
	\exp\left\{-\int\limits_0^{\bar{g}_s} \left[\frac{\tau_r(x)}{\beta_s(x)}-\frac{d_0}{b_0x} \right]\, dx \right\}\, .
\end{gather}
Here $\beta_s$ is the RG-function of the gauge coupling in the scheme $s$ introduced earlier (cf.~eq.~(\ref{eq:beta-function})),
while
\begin{equation}
	\tau_r(\bar{g}_s(\mu))={\rmd \ln \overline{m}_r(\mu)\over \rmd \ln\mu}\,,
\end{equation}
determines the scale dependence of the quark mass in the scheme $r$. At high energy, $\mu\to\infty$, the 
function $\tau_r(\bar{g}_s(\mu))$ admits a perturbative expansion in powers of $\bar{g}^2_s$, where the
leading order coefficient $d_0$ is scheme independent, i.e.
\begin{equation}
	\tau_r(\bar{g}_s)\overset{\bar{g}_s\to0}{\sim}-\bar{g}^2_s (d_0 + d^{r,s}_1 \bar{g}^2_s + \ldots)\,,
	\qquad
	d_0=6C_F/(4\pi)^2\,.
\end{equation}
From these definitions it is easy to show that the RGI mass $M$ is renormalization scheme and scale
independent and, thus, satisfies,
\begin{equation}
	M=\overline{m}_r(\mu)\varphi_{r,\rm m}(\bar{g}_s(\mu))=\overline{m}_{r'}(\mu')\varphi_{r',\rm m}(\bar{g}_{s'}(\mu'))\,.
\end{equation}
We moreover note that from eq.~(\ref{eq:M}), it is immediate to infer that the ratio of quark masses at
two different scales, $\mu_1,\mu_2$, can be written as
\begin{equation}
	\label{eq:massratio}
	{\overline{m}_{r}(\mu_2)\over \overline{m}_{r}(\mu_1)}
	=\exp\left\{\int\limits_{\bar{g}_s(\mu_1)}^{\bar{g}_s(\mu_2)} \frac{\tau_r(x)}{\beta_s(x)}\, dx \right\}\, .
\end{equation} 
Similarly to the determination of the $\Lambda$-parameter (cf.~Sects.~\ref{sec:LambdaParam} and \ref{sec:StepScaling}), 
also the ratio ${M/\overline{m}_{r}(\mu_{\rm  dec})}$ is obtained by breaking up the computation of the
RG running of the quark masses into two separate factors, each one referring to a different energy range,
\begin{equation}
	\label{eq:Momdec}
	{M\over \overline{m}_{r}(\mu_{\rm  dec})}= 
	{M\over \overline{m}_{r}(\mu_{0}/2)}\times
	{\overline{m}_{r}(\mu_{0}/2)\over \overline{m}_{r}(\mu_{\rm  dec})}\,,
\end{equation}
where $\mu_{0}= 4381(93)\, \MeV$ is the renormalization scale already introduced in Sect.~\ref{sec:StepScaling}. 

For the first term in eq.~(\ref{eq:Momdec}), we find~\cite{Campos:2018ahf},
\begin{equation}
	{M\over \overline{m}_{\rm SF}(\mu_{0}/2)}=1.7505(89)\,.
\end{equation}
The determination relies on finite-volume renormalization schemes for both the quark masses and the gauge coupling 
based on the Schr\"odinger functional (SF) of QCD. It is obtained by considering eq.~(\ref{eq:M}) and the 
following parameterizations for the RG-functions~\cite{Campos:2018ahf},
\begin{equation}
	\tau_{\rm SF}(\bar{g}_{\rm SF})=-\bar{g}_{\rm SF}^2\sum_{n=0}^{2}d_n\,\bar{g}_{\rm SF}^{2n}
	\qquad
	\beta_{\rm SF}(\bar{g}_{\rm SF})= -\bar{g}_{\rm SF}^3\sum_{n=0}^{3}b_n\,\bar{g}_{\rm SF}^{2n}
	\qquad
	\bar{g}_{\rm SF}^2\in[0,2.45] 
\end{equation}
where the coefficients $b_0,b_1,b_2$, and $d_0,d_1$ are fixed to their perturbative values,
with
\begin{gather}
	d^{\rm SF}_1={1\over(4\pi)^2} (0.2168 + 0.084\Nf )\,,\\ 
	b^{\rm SF}_2= {1\over(4\pi)^3} (0.483 - 0.275\Nf + 0.0361\Nf^2 - 0.00175\Nf^3)\,,
\end{gather}
scheme dependent, while 
\begin{equation}
	(4\pi)^3d^{\rm fit}_2=-0.18(52)\,,
	\qquad
	(4\pi)^4b^{\rm fit}_3=4(3)\,,
\end{equation}
were inferred from the data. The value of the gauge coupling corresponding to the scale $\mu=\mu_0/2$ 
is given instead by (remember that $\mu_0$ was defined by $\bar{g}^2_{\rm SF}(\mu_0)=2.012$)~\cite{DallaBrida:2018rfy},
\begin{equation}
	\bar{g}^2_{\rm SF}(\mu_0/2)=2.452(11)\,.
\end{equation}
For the second factor in eq.~(\ref{eq:Momdec}), we find,
\begin{equation}
	{\overline{m}_{\rm SF}(\mu_{0}/2)\over \overline{m}_{\rm SF}(\mu_{\rm  dec})}=0.8423(23).
\end{equation}
It is obtained considering eq.~(\ref{eq:massratio}) and the following parameterization 
of the ratio of RG-functions~\cite{Campos:2018ahf}
\begin{equation}
	\frac{\tau_{\rm SF}(\bar{g}_{\rm GF})}{\beta_{\rm GF}(\bar{g}_{\rm GF})} = 
	\frac{1}{\bar{g}_{\rm GF}}\sum_{n=0}^3 f_n \bar{g}_{\rm GF}^{2n}\,,
\end{equation}
with 
\begin{equation}
f_0=1.28493\,,
\quad
f_1=-0.292465\,, 
\quad
f_2=0.0606401\,,
\quad
f_3=-0.00291921\,. 	
\end{equation}
These parameters are correlated with covariance,
\begin{equation}
	\begin{split}
		{\rm cov}(f_i,f_j) =
		&\left(
		{\footnotesize
			\begin{array}{cccc}
				\phantom{+}2.33798\times 10^{-2} & -1.47011\times 10^{-2} &  \phantom{+}2.81966\times
				10^{-3} &
				-1.66404\times
				10^{-4} \\
				-1.47011\times 10^{-2} &  \phantom{+}9.54563\times 10^{-3} & -1.87752\times
				10^{-3} &
				\phantom{+}1.12962\times
				10^{-4} \\
				\phantom{+}2.81966\times 10^{-3} & -1.87752\times 10^{-3} &  \phantom{+}3.78680\times
				10^{-4} &
				-2.32927\times
				10^{-5} \\
				-1.66404\times 10^{-4} &  \phantom{+}1.12962\times 10^{-4} & -2.32927\times
				10^{-5} &
				\phantom{+}1.46553\times
				10^{-6} \\
			\end{array}
		}
		\right)\,.
	\end{split}
\end{equation}
The relevant values of the gauge coupling are given by (cf.~section \ref{sec:step-scaling-running} and \cite{DallaBrida:2016kgh,DallaBrida:2022eua}),
\begin{equation}
	\bar{g}^2_{\rm GF}(\mu_0/2) = 2.6723(64)\,,
	\qquad
	\bar{g}^2_{\rm GF}(\mu_{\rm dec}) = 3.949\,.	
\end{equation}
Combining the results above, we finally obtain,
\begin{equation}
	\label{eq:Movmbar}
	\frac{M}{\overline{m}_{\rm SF}(\mu_{\rm dec})} = 1.4744(85).
\end{equation}
We refer the interested reader to~\cite{DallaBrida:2022eua} and references therein 
for more details on the determinations presented in this section. 

We conclude by noticing that our analysis properly takes into account the 
uncertainties on $Z_{\rm m}$, $b_{\rm m}$, $am_{\rm crit}$ (and their correlation) as well as 
those on ${M}/\overline{m}_{\rm SF}(\mu_{\rm dec})$ in the determination of $z=M/\mu_{\rm  dec}$. 
These errors are propagated to the massive coupling $\bar{g}^2_{\rm GFT}(\mu_{\rm  dec},M)$
and finally to our result for $\Lambda^{(3)}_{\overline{\rm MS}}$ from decoupling.  
The uncertainty on $z$ contributes less than a $0.5\%$ to the final error squared of 
$\Lambda^{(3)}_{\overline{\rm MS}}$, eq.~(\ref{eq:LambdaNf3Final}).

\subsection{Boundary contributions to the decoupling limit}
\label{sec:boundary}

The GF coupling with SF boundary conditions employed in our decoupling strategy (cf.~eq.~(\ref{eq:GFTcoupl}))
is in principle affected by $\rmO(1/M)$ corrections to the large-mass limit. We can obtain an explicit 
expression for these contributions by invoking the tools of effective field theory. 

\subsubsection{Effective decoupling theory and $\rmO(1/M)$ terms}

Focusing on the continuum theory, at large quark mass, QCD with $\Nf$ mass-degenerate quarks can 
be described by an effective theory with action~\cite{Weinberg:1980wa,Athenodorou:2018wpk}, 
\begin{equation}
	\label{eq:Sdec}
	S_\text{dec} = S_{0,\text{dec}} + \frac{1}{m} S_{1,\text{dec}}+\frac{1}{m^2} S_{2,\text{dec}} + \ldots\,,
\end{equation}
where for the time being $m$ refers to a generic quark mass to be specified later.
Since all quarks  decouple  simultaneously, the leading term $S_{0,\text{dec}}$ corresponds to the pure gauge action,
\begin{equation}
	S_{0,\text{dec}} = 
	-\frac{1}{2g^2} \int d^4x\, \tr(F_{\mu\nu} F_{\mu\nu})\,,
\end{equation}
where $F_{\mu\nu}$ is the Yang-Mills field-strength tensor and $g$ a yet unspecified gauge coupling. 
The other terms, $S_{k,\text{dec}}$ with $k>0$, are  given by space-time integrals 
of gauge invariant local operators, polynomial in the gauge field and its derivatives. 
In the case of a theory with no space-time boundary these are of mass dimension $4+k$.
Moreover, gauge invariance and $O(4)$ symmetry do not allow for odd values of $k$, so 
that the $S_{1,\rm dec}$ term must vanish. 

Different is the situation in the presence of SF boundary conditions~\cite{Luscher:1992an,Sint:1993un}. In this case, additional terms are 
allowed in the effective action (\ref{eq:Sdec}). The $S_{k,\rm dec}$ terms can in fact include also space 
integrals of dimension $3+k$ operators localized at $x_0=0,T$. Considering the homogeneous SF boundary conditions
employed in our strategy and the symmetries of the resulting theory, one can show that we can now have a linear 
term in $1/m$ of the form~\cite{DallaBrida:2022eua}
\begin{equation}
	S_{1,\rm dec}= \int d^3x\,\omega_{{\rm b}}(g)
	\big[\mathcal{O}_{{\rm b}}(0,\boldsymbol{x})+  \mathcal{O}_{{\rm b}}(T,\boldsymbol{x})\big]\,,
	\quad
	\mathcal{O}_{\rm b}(x)= -{1\over g^2}\sum_{k=1}^3\tr(F_{0k}(x)F_{0k}(x))\,.
\end{equation}
In this equation, $\omega_{\rm b}(g)$ is a coefficient function that needs to be 
properly adjusted in order to match the results of QCD with $\Nf$ heavy quarks and those of 
the effective theory up to $\rmO(1/m^2)$ corrections. Its value has been computed at 1-loop 
order in perturbation theory in ref.~\cite{DallaBrida:2022eua}, and it is given by%
\footnote{Clearly, the value of the leading order coefficient $\omega_{{\rm b}}^{(1)}$
does not depend on the choice of coupling $g$, which is yet unspecified.}
\begin{equation}
	\label{eq:omegaLO}
	\omega_{\rm b}(g) = 
	\omega_{{\rm b}}^{(1)} g^2 + \omega_{{\rm b}}^{(2)} g^4+ \ldots\,,
	\qquad
	\omega_{{\rm b}}^{(1)}=-0.0541(5) N_{\rm f}/(4\pi)\,.
\end{equation}
In the case of the GF coupling, the knowledge of the effective action is a crucial requirement
to estimate the $\rmO(1/m)$ corrections to its large-mass limit. Following the discussion
in ref.~\cite{DallaBrida:2022eua}, we can in fact write,
\begin{multline}
	\label{eq:LeadingO1MCorr}
	\bar{g}^2_{\rm GFT}(\mu,M) \overset{M\to\infty}{=} \big[\bar{g}^{(0)}_{\rm GFT}(\mu)\big]^2-\\[2ex]
	+{\omega_{\rm b}(g_\star)\over m_\star }
	\int d^3x\,
	\bigg\langle {t^2E(t,y)\over \mathcal{N}}\big[\mathcal{O}_{{\rm b}}(0,\boldsymbol{x})+  \mathcal{O}_{{\rm b}}(T,\boldsymbol{x})\big]\bigg\rangle^{\rm conn}_{\rm YM}
	+\rmO\bigg({1\over m_\star^2}\bigg)\,.
\end{multline}
In the above equation $\langle\cdots\rangle_{\rm YM}^{\rm conn}$ stands for the connected expectation value in the Yang-Mills theory
with action $S_{0,\rm dec}$, in the presence of homogeneous SF boundary conditions, and geometry $T=2L$. 
The coupling $g_\star=\bar{g}_{\overline{\rm MS}}(m_\star)$ and quark mass $m_\star=\overline{m}_{\overline{\rm MS}}(m_\star)$
are those of the $\Nf$-flavour theory, while the energy density $E(t,y)$ is evaluated for 
$\sqrt{8t}=cL$ and $y_0=T/2$ (cf.~Sect. \ref{sec:cont-limit-mass}).

\subsubsection{Leading order estimate of the $\rmO(1/M)$ corrections}

We can obtain an estimate for the size of the $\rmO(1/m)$ term in eq.~(\ref{eq:LeadingO1MCorr}) 
using lattice simulations~\cite{DallaBrida:2022eua}. To this end, we define the difference,
\begin{equation}
	\Delta(z)\equiv \bar{g}^2_{\rm GFT}(\mu_{\rm dec},M)-[\bar{g}^{(0)}_{\rm GFT}(\mu_{\rm dec})]^2\,,
	\qquad
	z=M/\mu_{\rm dec}\,.
\end{equation}
Given eq.~(\ref{eq:LeadingO1MCorr}) and the leading order (LO) result, eq.~(\ref{eq:omegaLO}), for 
the matching coefficient $\omega_{\rm b}$, we can write the LO estimate for the $\rmO(1/M)$ corrections 
to the massive GFT coupling as,
\begin{equation}
	\label{eq:MasterEq}
	\Delta(z)\big|_{\rm LO}
	\approx{\omega_{\rm b}^{(1)}g_\star^2 \over z}\bigg({M\over m_\star}\bigg)
	p_1\big(\bar{g}^2_{\rm GFT}(\mu_{\rm dec},M)\big)
	+ {\rm O}\bigg(g_\star^4,{1\over z^2}\bigg)\,.
\end{equation}
Because $g_\star\to0$ for $M\to\infty$, this estimate becomes more accurate the larger $M$ is,
since eq.~(\ref{eq:omegaLO}) becomes a better approximation to $\omega_{\rm b}$. 
The coefficient function $p_1$ is given by the matrix element,
\begin{equation}
	\label{eq:p1}
	p_1\equiv 
	-\lim_{a\to0}
	\int d^3x\,
	\bigg\langle{t^2E(t,y)\over \mathcal{N}}L\big[\mathcal{O}_{{\rm b}}(0,\boldsymbol{x})+  
	\mathcal{O}_{{\rm b}}(T,\boldsymbol{x})\big]\bigg\rangle^{\rm conn}_{\rm YM}\,.
\end{equation}
This must be computed, non-perturbatively, along a line of constant physics defined by 
a fixed value of $\bar{g}^{(0)}_{\rm GFT}(\mu_{\rm dec})\equiv\bar{g}_{\rm GFT}(\mu_{\rm dec},M)$. 

\begin{table}[hpbt]
	\centering
	\begin{tabular}{cccc}
		\toprule
		$c$ & $\bar{g}^2_{\rm GFT}(\mu_{\rm dec},M)$ & $L/a$ & 
		$p_1$ \\
		\midrule
		\multirow{4}{*}{$0.3$} &
		\multirow{4}{*}{$4.448$} 
		&   $10$ & $-0.396(62)$\\
		& & $12$ & $-0.370(73)$\\
		& & $16$ & $-0.305(49)$\\
		& & $\infty$ & $-0.15(17)$ \\	    
		\midrule
		\multirow{4}{*}{$0.36$} &
		\multirow{4}{*}{$5.347$} 
		& $10$ & $-0.94(10)$\\
		& & $12$ & $-0.86(12)$\\
		& & $16$ & $-0.77(07)$\\
		& & $\infty$ & $-0.48(26)$ \\	    
		\midrule	
		\multirow{4}{*}{$0.42$} &
		\multirow{4}{*}{$6.690$} 
		& $10$ & $-2.06(17)$ 	\\
		& & $12$ & $-1.88(20)$	\\ 
		& & $16$ & $-1.71(12)$ 	\\   
		& & $\infty$ & $-1.14(43)$ \\	    
		\bottomrule
	\end{tabular}
	\caption{Results for $p_1$ at finite lattice spacing. 
		Estimates of $a/L\to0$
		extrapolations linear in $a/L$ are also given.}
	\label{tab:dgsqdct_cont}
\end{table}

For the lattice computation of $p_1$ we followed the strategy described in ref.~\cite{DallaBrida:2022eua}. 
We refer the interested reader to this reference for the details on the lattice set-up, 
and in particular for a discussion on the lattice discretization of the relevant operators. 
We performed simulations for $L/a=10,12,16$, at $3$ or $4$ values of $\beta$ depending on the value of
$L/a$. The values of $\beta$ were chosen in such a way that $p_1$ can be interpolated in $\beta$ so that 
the continuum limit can be taken at fixed $\bar{g}^{(0)}_{\rm GFT}(\mu_{\rm dec})$ corresponding to $\bar{g}_{\rm GFT}(\mu_{\rm dec},M)$ with 
$z=6$ and $c=0.3,0.36,0.4$. The results for $p_1$ we obtained are collected in Table \ref{tab:dgsqdct_cont} 
for the 3 different lattice resolutions. We also provide an estimate of the continuum limit 
from linear extrapolations in $a/L$.

Given the non-perturbative results for $p_1$, all that 
is needed to compute $\Delta(z)|_{\rm LO}$
are the values for $g_\star$ and $M/m_\star$
at the relevant quark mass. The latter can be obtained
once 
\begin{equation}
	{M\over\Lambda^{(3)}_{\overline{\rm MS}}}
	=z\times {\mu_{\rm dec}\over\Lambda^{(3)}_{\overline{\rm MS}}}\,,
\end{equation}
is specified, thanks to the perturbative knowledge of the RG functions in the $\overline{\rm MS}$ scheme~\cite{Bernreuther:1981sg,Grozin:2011nk,Chetyrkin:2005ia,Schroder:2005hy,Kniehl:2006bg,Gerlach:2018hen,vanRitbergen:1997va,Czakon:2004bu,Baikov:2016tgj,Herzog:2017ohr,Luthe:2017ttg,Chetyrkin:2017bjc}. 
Taking ${\mu_{\rm dec}/\Lambda^{(3)}_{\overline{\rm MS}}}\approx 2.3$ 
and $z=6$, we find for $\Nf=3$,
\begin{equation}
	g^2_\star\approx 3.0277\,,
	\qquad
	M/m_\star\approx  1.4889\,,
\end{equation}
where the 5-loop $\beta_{\overline{\rm MS}}$ and 4-loop $\tau_{\overline{\rm MS}}$ functions were used.

\begin{table}[hpbt]
	\centering
	\begin{tabular}{ccc}
		\toprule
		$c$ & $\bar{g}^2_{\rm GFT}(\mu_{\rm dec},M)$ & 
		$\Delta(z=6)|_{\rm LO}$ \\
		\midrule
		$0.30$ & $4.448(14)$ & $0.0029$ \\	    
		$0.36$ & $5.347(22)$ & $0.0075$ \\	    
		$0.42$ & $6.690(37)$ & $0.0170$ \\	    
		\bottomrule
	\end{tabular}
	\caption{Results for $\Delta(z)|_{\rm LO}$ for $z=6$ and different values of $c$.}
	\label{tab:Deltaz}
\end{table}

In order to obtain a generous estimate for $\Delta(z)|_{\rm LO}$ we consider the value of $p_1$ at $L/a=16$. This gives 
a more conservative choice than taking the continuum limit values (cf.~Table \ref{tab:dgsqdct_cont}). 
The results for  $\Delta(z)|_{\rm LO}$ computed in this way are collected in Table \ref{tab:Deltaz}.
Comparing the results for $\Delta(z)|_{\rm LO}$ with the statistical uncertainties
on the corresponding massive coupling $\bar{g}^2_{\rm GFT}(\mu_{\rm dec},M)$, we see that the former
range from being about a factor 4-5 smaller for $c=0.30$, to $\approx3$ times smaller 
for $c=0.36$, to $\approx 2$ times smaller for $c=0.42$. We conclude that these effects can 
be safely discarded for our preferred choice, $c=0.36$
and $z=6$. They decrease further $\sim1/z$ as we reach the 
decoupling limit.

\subsection{Massive gradient flow coupling in perturbation theory} 
\label{sec:1loopMassiveGFcoupling}

It is instructive to study the decoupling of heavy quarks at the lowest non-trivial order in perturbation theory. The discussion will give us
some qualitative insight on the size of the leading corrections to the infinite quark-mass limit in computing the $\Lambda$-parameter via the decoupling method. 
Concomitantly, we hope that it will help the reader to better understand our strategy in simpler, perturbative, terms.

Unfortunately, there are no known perturbative results in QCD for the specific coupling definition we employed in our strategy, i.e.~the GF coupling defined in a finite volume with SF boundary conditions; only the pure gauge part has been obtained so far~\cite{DallaBrida:2017tru}. There are, however, results for an infinite space-time volume~\cite{Luscher:2010iy,Harlander:2016vzb}. In the following we shall consider these in order to get a qualitative picture for our case. We note that in the limit where $c=\sqrt{8t}/L\to0$, our coupling tends to the infinite volume one with corrections of O($c^4$). 

\subsubsection{One-loop GF coupling in infinite volume}

We begin by considering the perturbative results of ref.~\cite{Harlander:2016vzb}
for the energy density at positive flow time $\langle t^2E(t)\rangle$
(cf.~eq.~(36) of that reference). After a proper normalization by an overall
constant, we readily obtain the perturbative expansion of the GF
coupling at the 1-loop order and for finite quark masses, 
\begin{gather}
	\label{eq:GFcoupling1lp}
	\bar{g}^2_{\rm GF}(\mu,\overline{m}(\mu))=	
	\bar{g}^2_s(\mu) +c_1(z) \bar{g}^4_s(\mu) + \rmO(\bar{g}_s^6)\,,\\[1ex]
	c_1(z)=(1.0978 + 0.0075\,\Nf)/(4\pi) - \frac{\Nf}{24\pi^2} \Omega_{1}(z)\,,
\end{gather}
where $\mu=1/\sqrt{8t}$, $\bar{g}_s^2\equiv\bar{g}^2_{\overline{\rm MS}}(\mu)$, and
$z\equiv\overline{m}(\mu)/\mu$. Note that we restrict to the case of $\Nf$ mass-degenerate 
quarks with renormalized quark mass $\overline{m}(\mu)$. At this order in perturbation
theory there is no need to specify a renormalization scheme for 
the mass, which is therefore left unspecified. 

The function $\Omega_1(z)$, which encodes the mass dependence of the coupling, is given here for completeness,
\begin{equation}
	\label{eq:omega1q}
	\Omega_{1}(z) = 
	1 - \gamma_{\rm E} - \ln (z^2/4) - z^2 + z^2 I(z)\,,
\end{equation}
where
\begin{equation}
	\begin{split}
	I(z) = \frac12\int_0^\infty\rmd x\,e^{-x/4}\left (1+\frac{x}{4z^2} \right )\left(1-\frac{x}{2z^2} \right) 
	\frac{u\ln u}{u^2-1}\,,\\[1.5ex]
	u = \frac{\sqrt{x+4z^2}-\sqrt x}{\sqrt{x+4z^2}+\sqrt x} \,.
	\end{split}
\end{equation}
In the following, we will mostly be interested in its limits for small and large
values of $z$, 
\begin{equation}
	\label{eq:omega1qlimits}
	\Omega_{1}(z)\quad\to\quad
	\left\{
	\begin{array}{l}
		-{3\over2}\, z^2 + \rmO({z^4})\,,\\[2.5ex]
		-2\ln z - h -
		{8\over 5z^2} + \rmO({z^{-4}})\,,
	\end{array}
	\right.
\end{equation}
with $h=\gamma_{\rm E} + {2\over3} - 2\ln2$.

\subsubsection{Renormalization group improved coupling in the one-loop model}
\Eq{eq:GFcoupling1lp} is a well behaved perturbative expansion for 
small $z$, where the coefficient function $c_1(z)$ remains finite.
However, $c_1(z)$ diverges in the limit of large $z$ and the expansion becomes useless. Renormalization group (RG) improvement 
removes such an issue (and has already been silently used,
setting $\mu=1/\sqrt{8t}$), but for large $z$ a second scale 
appears and RG improvement is not generally
applicable to our knowledge. One could  resort to the effective
theory at large $z$ at the expense of having to patch up domains of small and large $z$. Instead, here, we give a full RG improved expression valid for all $z$. Since it employs a non-systematic approximation, we call this the one-loop model.

We start from the truncated RG equations
\begin{eqnarray}
\label{eq:RGEg}
	\mu{\rmd\over \rmd\mu} {1\over \bar{g}^2_{\rm GF}}&=&
	2b_0(\Nf) + \frac{\Nf}{24\pi^2} 
	\bigg(\mu {\rmd\over \rmd\mu}\bigg)\Omega_1(z)\,,	
	\\
\label{eq:RGEm}
		\mu{\rmd\over \rmd\mu} \mbar  &=& 0		\,,
\end{eqnarray}
where $b_0(\Nf)$ is the universal lowest order coefficient of 
the perturbative expansion of the $\beta$-function for $\Nf$-flavours. While in \eq{eq:RGEg} the one-loop term is taken along, in \eq{eq:RGEm} it is dropped. It only contributes via $c_1$ in \eq{eq:GFcoupling1lp} and therefore is higher order, as we have remarked before. However, even in the massless case, one has to include the two-loop term in the $\beta$-function and the one-loop term in the mass anomalous dimension in order to get a systematic expansion of the $\Lambda$-parameter and the RGI mass in terms of the coupling, see eqs.~(\ref{eq:varphi}, \ref{e:Lambda_from_mu}) and (\ref{eq:varphim}, \ref{eq:M}). We do not do this here and therefore we use the term model, not approximation.

The truncated RG equations can be integrated exactly to obtain,
\begin{equation}
	{1\over \bar{g}^2_{\rm GF}}=
	2b_0(\Nf)\ln\big({\mu/\Lambda_{\rm GF}^{(\Nf)}}\big)
	+{\Nf\over 24\pi^2}\Omega_1(z)\,, \quad \mbar(\mu) = M,
\end{equation}
where $\Lambda_{\rm GF}^{(\Nf)}$ is a conveniently chosen 
constant; taking $z\to0$ (and $b_1=d_0=0$ as fit for the model), we see that it is the $\Lambda$-parameter in the GF scheme for the $\Nf$-flavour theory. In the same way, the constant $M$ can be identified 
with the RGI mass or the running mass.

At large values of $z$, the quarks decouple. We thus expect that,
\begin{equation}
	{1\over \bar{g}^2_{\rm GF}}
	\overset{z\to\infty}{=}
	2b_0(0)\ln\big({\mu/\Lambda_{\rm GF, dec}^{(0)}}\big)
	+\rmO(1/z^2)\,,
\end{equation}
for a properly chosen function $\Lambda^{(0)}_{\rm GF, dec}\equiv\Lambda^{(0)}_{\rm GF, dec}(\Lambda_{\rm GF}^{(\Nf)},M)$
of the $\Lambda$-parameter of the $\Nf$-flavour theory, $\Lambda_{\rm GF}^{(\Nf)}$ , and of the RGI quark mass $M$. 
We call $\Lambda^{(0)}_{\rm GF, dec}$ the $\Lambda$-parameter of the effective theory.

Taking into account the asymptotic limit of $\Omega_1(z)$ for large $z$ (cf.~eq.~(\ref{eq:omega1qlimits})),
\begin{equation}
	\Omega_1(z)= -2\ln(z) - h + \Omega_1^{\rm sub}(z)\,,
	\quad 
	\Omega_1^{\rm sub}(z)=\rmO({1/z^{2}})\,,
\end{equation}
we find
\begin{equation}
	\label{eq:Lambda_dec_asy}
	{1\over \bar{g}^2_{\rm GF}}
	\overset{z\to\infty}{=}
	2b_0(\Nf)\ln(\mu/\Lambda_{\rm GF}^{(\Nf)}) - 
	{\Nf\over 24\pi^2}(2\ln(z) +h )+\rmO(1/z^2)\, 
	\equiv
	2b_0(0) \ln(\mu/\Lambda^{(0)}_{\rm GF, dec})\,,
\end{equation}
and with
\begin{equation}
	b_0(\Nf) = b_0(0)-{\Nf\over 24\pi^2} \,,
\end{equation}
we infer the relation between the $\Lambda$-parameter of the $\Nf$-flavour theory and
that of the effective theory, 
\begin{equation}
	\label{eq:PGF1lp}
	P_{\rm GF}\equiv
	{\Lambda^{(0)}_{\rm GF,dec}\over\Lambda_{\rm GF}^{(\Nf)}} = 
	\bigg({M\over \Lambda_{\rm GF}^{(\Nf)}}\bigg)^{\eta_0(\Nf)} 
	\exp\bigg({h\eta_0(\Nf)\over 2}\bigg)\,, 
	\quad 
	\eta_0(\Nf)=1-{b_0(\Nf)\over b_0(0)}\,.
\end{equation}
This relation can also be obtained from the 
ratio of $\Lambda$-parameters in the $\overline{\rm MS}$-scheme,
i.e. the function $P$ in eq.~(\ref{eq:Pdef}), setting $b_i=d_{i-1}=0,\,i>0$ and 
changing  to the GF scheme for both 
$\Lambda^{(0)}$ and $\Lambda^{(3)}$. Note that this scheme-change is a $z$-independent factor. 

\subsubsection{Leading $\rmO(1/z^2)$ corrections to the decoupling relation in the model}

We can gain some insight into the leading order corrections
to the decoupling relation by keeping the $\rmO(1/z^2)$ terms
in the large $z$ expansion of the function $\Omega_{1}(z)$
when establishing the relation eq.~(\ref{eq:Lambda_dec_asy}). 
Along the lines of our decoupling strategy 
we thus introduce effective $\Lambda$-parameters (cf.~section~\ref{sec:LambdaFromDec}), 
$\Lambda^{\rm eff}\equiv \Lambda_{\rm GF}^{(\Nf),\rm eff}$ and
$\Lambda_{\rm dec}^{\rm eff}\equiv\Lambda^{(0),\rm eff}_{\rm GF, dec}$,
for the $\Nf$-flavour and effective theory respectively, through the definitions
\begin{equation}
	\begin{split}
	{1\over \bar{g}^2_{\rm GF}}
	&\hspace*{2.25mm}{=}\hspace*{1.75mm}
	2b_0(\Nf)\ln(\mu/\Lambda) - 
	{\Nf\over 24\pi^2}(2\ln(z) +h-\Omega_1^{\rm sub}(z))
	\,,\\[1ex]
	&\hspace*{2.25mm}{\equiv}\hspace*{1.75mm}
	2b_0(\Nf)\ln(\mu/\Lambda^{\rm eff}) - 
	{\Nf\over 24\pi^2}(2\ln(z) +h)
	\,,\\[2.75ex]
	&\hspace*{2.25mm}{\equiv}\hspace*{1.75mm}
	2b_0(0) \ln(\mu/\Lambda^{\rm eff}_{\rm dec})\,,
	\end{split}
	\label{eq:Lambdaeff}
\end{equation}
where for the ease of notation we also define $\Lambda\equiv\Lambda_{\rm GF}^{(\Nf)}$  
and $\Lambda_{\rm dec}\equiv\Lambda^{(0)}_{\rm GF, dec}$. First note that,
analogously to $\Lambda$ and $\Lambda_{\rm dec}$,  the effective $\Lambda$-parameters are
related by (cf.~eqs.~(\ref{eq:Lambda_dec_asy})-(\ref{eq:PGF1lp}))%
\footnote{In our strategy the $\Lambda$-parameters are first converted to the $\overline{\rm MS}$-scheme 
and the function $P_{\overline{\rm MS}}$ is used. However, since the ratio $P_\mathrm{GF}/P_\msbar$ is 
independent of $z$, this change of schemes is irrelevant.}
\begin{equation}
	\label{eq:RatioLambdaeff}
 	\Lambda_{\rm dec}^{\rm eff} =
 	P_{\rm GF}(M/{\Lambda^{\rm eff}})
 	{\Lambda^{\rm eff}}=
   (M/\Lambda^{\rm eff})^{\eta_0}
   	e^{{h\eta_0\over 2}}
   	{\Lambda^{\rm eff}}\,.
\end{equation}
Second, by construction, in the limit $z\to\infty$, $\Lambda^{\rm eff}$ and $\Lambda^{\rm eff}_{\rm dec}$ 
approach their respective counterparts $\Lambda$ and $\Lambda_{\rm dec}$ with corrections of $\rmO(1/z^2)$. 
In particular, from the definitions in eq.~(\ref{eq:Lambdaeff}), we have that,
\begin{equation}
	\label{eq:CorrRatioLambdaeff}
	\ln(\Lambda^{\rm eff}/\Lambda)
	= -\frac{\eta_0(\Nf)}{2-2\eta_0(\Nf)} \Omega_1^{\rm sub}(z)\,.
\end{equation}
Considering the large $z$ expansion of $\Omega_1(z)$ in eq.~(\ref{eq:omega1qlimits}),  
this gives 
\begin{equation}
	{\Lambda^{\rm eff}\over\Lambda}
	= 1 + \frac{\eta_0(\Nf)}{2-2\eta_0(\Nf)} \frac{8}{5z^2} + \rmO({1/z^4})
	\overset{\Nf=3}{=}
	1+ \frac{8}{45z^2} + \rmO({1/z^4})\,,
	\label{eq:CorrRatioLambdaEff2}
\end{equation}
where the coefficient of the neglected $1/z^4$ term is of a similar size. 

Considering that in our finite volume set-up $z=M/\mu=cML$, we find that, for $\Nf=3$, 
the above correction term is around 4\% for $c=0.36$ and $ML=6$. 
Instead, if we use for $M$ in the model  the value of $\mbar(\mu)$ in the 
non-perturbative computation and use the non-perturbative relation 
$M/\overline{m}_{\rm SF}(\mu_{\rm  dec})\approx1.47$, eq.~(\ref{eq:Movmbar}), the
effect we find is about two times larger. This level of ambiguity should not be surprising given the model used 
in our calculation; we recall that the coupling is computed only at one-loop order in perturbation 
theory (in infinite volume), and the running of the quark mass is neglected all-together.
Nonetheless, the model captures the right bulk part of the effect. 
Looking at figure \ref{fig:mtoinf}, that shows our non-perturbative results for $\Lambda^{\rm eff}$,
we see that for $c=0.36$ and $ML=6$ the correction is about 6\% with respect to the infinite mass limit.

The result in eq.~(\ref{eq:CorrRatioLambdaEff2}) also elucidates how the approach to the infinite mass 
limit can be tested when working at fixed $L$ by considering different values of $c$. In particular, 
larger values of $c$ are expected to lead to smaller corrections to the decoupling limit as they 
effectively correspond to larger ratios $M/\mu$.%
\footnote{We recall that, instead, statistical uncertainties in $\bar{g}^2_{\rm GF}$ and thus $\Lambda^{\rm eff}$ tend 
to increase for larger $c$-values.}
Following this reasoning, our final determination of $\Lambda$ is based on  the results 
for $c=0.36$,  which strikes a good balance between systematic and statistical uncertainties; both of them are small.

\subsubsection{Illustration}
The behavior of the coupling in the one loop model is shown in
\fig{fig:cont} (left) together with the massless coupling and the
one in the decoupled theory with $\Lambda^{(0)}_{\rm GF,dec}$ from \eq{eq:PGF1lp}. The figure further illustrates the trajectory that one follows in the non-perturbative decoupling strategy.

\subsection*{Acknowledgments}

We wish to express our gratitude to Martin L\"uscher, Peter Weisz and
Ulli Wolff, whose work on finite size scaling for renormalized
couplings, the Schr\"odinger Functional and the gradient flow laid the
foundations for the results presented here. We also thank them for
numerous enlightening discussions and advice over the years.
Furthermore, we are indebted to our colleagues in the ALPHA
collaboration, especially Simon Kuberski, Patrick Fritzsch, Jochen
Heitger and Hubert Simma for sharing their data on the renormalized
coupling.
We thank Giulia Zanderighi for sharing the data that enters
in Figure~\ref{fig:alphas}.

The work is supported by the Spanish MCIU project No. CNS2022-136005,
the Generalitat Valenciana genT program No. 
CIDEGENT/2019/040, the German Research Foundation (DFG) research unit
FOR5269 “Future methods for studying confined gluons in QCD”. 

AR acknowledges financial support by the European projects
H2020-MSCA-ITN-2019//860881-HIDDeN and 101086085-ASYMMETRY, and the
national project PID2023-148162NB-C21.  
RH was supported by the programme “Netzwerke 2021”, an initiative of
the Ministry of Culture and Science of the State of Northrhine
Westphalia, in the NRW-FAIR network, funding code NW21-024-A. SS and
RS acknowledge funding by the H2020 program in the Europlex training
network, grant agreement No. 813942. Generous computing resources were
supplied by the North-German Supercomputing Alliance (HLRN, project
bep00072) and by the John von Neumann Institute for Computing (NIC) at
DESY, Zeuthen, and the local SOM clusters, funded by the
MCIU with funding from the European Union NextGenerationEU
(PRTR-C17.I01) and Generalitat Valenciana Grant No. ASFAE/2022/020.


\putbib  
\end{bibunit}

\end{document}